\documentclass[prd,a4paper,superscriptaddress,nofootinbib]{revtex4}
\usepackage{graphicx}
\usepackage{color}

\extrafloats{100}

\newcommand{\be}{\begin{equation}}
\newcommand{\ee}{\end{equation}}
\newcommand{\bq}{\begin{eqnarray}}
\newcommand{\eq}{\end{eqnarray}}
\newcommand{\bsq}{\begin{subequations}}
\newcommand{\esq}{\end{subequations}}
\newcommand{\bc}{\begin{center}}
\newcommand{\ec}{\end{center}}
\begin{document}
\title{Evolution of Semilocal String Networks: II. Velocity estimators}
\author{A. Lopez-Eiguren}
\email{asier.lopez@ehu.es}
\affiliation{Department of Theoretical Physics, University of the Basque Country UPV-EHU, 48040 Bilbao, Spain}
\author{J. Urrestilla}
\email{jon.urrestilla@ehu.es}
\affiliation{Department of Theoretical Physics, University of the Basque Country UPV-EHU, 48040 Bilbao, Spain}
\author{A. Ach\'ucarro}
\email{achucar@lorentz.leidenuniv.nl}
\affiliation{Institute Lorentz of Theoretical Physics, University of Leiden, 2333CA Leiden, The Netherlands}
\affiliation{Department of Theoretical Physics, University of the Basque Country UPV-EHU, 48040 Bilbao, Spain}
\author{A. Avgoustidis}
\email{anastasios.avgoustidis@nottingham.ac.uk}
\affiliation{School of Physics and Astronomy, University of Nottingham, University Park, Nottingham NG7 2RD, England}
\author{C. J. A. P. Martins}
\email{Carlos.Martins@astro.up.pt}
\affiliation{Centro de Astrof\'{\i}sica da Universidade do Porto, Rua das Estrelas, 4150-762 Porto, Portugal}
\affiliation{Instituto de Astrof\'{\i}sica e Ci\^encias do Espa\c co, CAUP, Rua das Estrelas, 4150-762 Porto, Portugal}

\date\today
\begin{abstract}
We continue a comprehensive numerical study of semilocal string
networks and their cosmological evolution. These can be thought of as
hybrid networks comprised of (non-topological) string segments, whose
core structure is similar to that of Abelian Higgs vortices, and whose
ends have long--range interactions and behaviour similar to that of
global monopoles. Our study provides further evidence of a linear scaling regime, already
reported in previous studies, for the typical length scale and velocity
of the network. We introduce a new algorithm to identify the position
of the segment cores. This allows us to determine the length and
velocity of each individual segment and follow their evolution in
time. We study the statistical distribution of segment lengths and
velocities for radiation- and matter-dominated evolution in the regime
where the strings are stable. Our segment detection algorithm gives
higher length values than previous studies based on indirect detection
methods. The statistical
distribution shows no evidence of (anti)correlation between the speed
and the length of the segments.
\end{abstract}
\maketitle

\section{Introduction}
\label{intro}

Topological defect networks \cite{Kibble:1980mv,Hindmarsh:1994re,Vilenkin:2000jqa,Copeland:2009ga} are predicted to form in high-energy physics models of the early universe \cite{Vilenkin:2000jqa,Jeannerot:2003qv,Dvali:1998pa}. In many cases they are stable enough to survive throught the subsequent expansion history of the universe, as fossil relics of its earlier stages. Advances in observational astrophysics and cosmology including sensitive observations of the cosmic microwave background \cite{Adam:2015rua,Hanson:2013hsb,Story:2012wx,Naess:2014wtr,Ade:2014afa,Ade:2014xna}, pulsar timing arrays \cite{Manchester:2012za,Kramer:2013kea} and more recently the detection of gravitational waves \cite{Abbott:2016blz,Abbott:2016nmj} provide exciting opportunities to characterize or at least constrain the physics of the early universe through cosmic strings.

Among the various types of possible defect networks, string-type ones are the most common (their formation being inevitable in many models) as well as the most interesting since (unlike other defects, such as monopoles and domain walls) their evolution is broadly speaking cosmologically benign. Defects of this type include the simplest Nielsen-Olesen line-like topological defects in field theories breaking $U(1)$ symmetry \cite{Nielsen:1973cs}, coherent macroscopic states of fundamental superstrings (known as F-strings), and D-branes extended in one macroscopic direction (D-strings). The latter two examples, collectively referred to as cosmic superstrings \cite{Polchinski:2004ia}, are generically predicted in string theoretic inflationary models involving spacetime-wrapping D-branes \cite{Dvali:1998pa,Burgess:2001fx,Majumdar:2002hy,Copeland:2003bj}. Various other hybrid networks are also possible, including the semilocal strings \cite{Vachaspati:1991dz,Achucarro:1999it} that are the focus of this work.

Understanding the evolution of these defect networks, and in particular the distinguishing features of each specific model class, is essential in order to fully exploit the forthcoming data. Since these are intrinsically non-linear objects, the quantitatively accurate modeling of defect network evolution is a difficult problem: analytic models can be obtained (using reasonable approximations), but any such model will necessarily contain some free parameters which cannot be calculated {\it ab initio}, but must be inferred by direct comparison with sufficiently high resolution numerical simulations---thereby calibrating the analytic model. The canonical approach to this modelling process is embodied in the Velocity-dependent One-Scale model (hereafter VOS model) and its various extensions \cite{Martins:1996jp,Martins:2000cs,Moore:2001px,Martins:2008zz,Avgoustidis:2007aa,Martins:2016ois}. Naturally, the complexity of the model is directly related to the complexity of the underlying particle physics model.

In the simplest scenario, corresponding to the Nielsen-Olesen solutions of the Abelian Higgs (hereafter AH) model there is clear analytic and numerical evidence that in simple radiation or matter dominated universes the network will reach a self-similar scaling regime, characterised by a single length-scale---the correlation length---which asymptotes to a constant fraction of the horizon. This allows us to use numerical simulations of limited dynamical range to infer properties of string networks evolving in the real Universe over much larger timescales. In practice, in a realistic universe containing both radiation, matter and dark energy, a cosmic string network never reaches this asymptotic scaling regime \cite{Azevedo:2017xev}, but it approaches it relatively fast. The situation is less clear for non-Abelian strings, as well as for cosmic superstring networks, despite much recent activity in this area.

Semilocal strings are a hybrid type of defect arising in theories with both local and global symmetries \cite{Vachaspati:1991dz,Achucarro:1999it}. For example the standard semilocal model is a minimal extension of the Abelian Higgs model by a global SU(2) symmetry. This model, which has an SU(2) doublet of two equally charged Higgs fields under a single U(1) gauge field, admits stable string solutions even though the vacuum manifold is simply connected. They are also well-motivated from the theoretical point of view, arising in supersymmetric grand unified theories of inflation \cite{Urrestilla:2004eh} and the corresponding D3/D7 brane inflation models \cite{Dasgupta:2004dw}. These are a natural extension of usual inflationary models, in which the only extra ingredient is the doubling of a hypermultiplet. Being non-topological, semilocal strings posess remarkably different microphysical properties, as compared to their topological counterparts. The most noteworthy of these is that they form finite open segments whose ends have long-range interactions akin to global monopoles \cite{Hindmarsh:1992yy}.

The cosmological evolution of semilocal networks is correspondingly distinctive. Their dynamics is dominated by the long range forces between the monopoles which imply that the individual segments can shrink and annihilate or grow by joining with other segments. The analytic modeling of these networks in a VOS-like context was started in \cite{Nunes:2011sf}, based on the premise that they can be described as local strings ending on global monopoles\footnote{The VOS model for global monopoles has been recently revisited in \cite{Sousa:2017wvx}}. In addition to describing the overall evolution of the network, this also includes a phenomenological description of the evolution of individual semilocal segments and a preliminary comparison to numerical simulations, though all these were somewhat limited by the lack of numerical simulations with adequate resolution.

More recently, we have taken advantage of progress in computing power and started a more systematic study of the cosmological evolution of semilocal string networks, by presenting the first detailed numerical study of these networks \cite{Achucarro:2013mga}, exploring a broad range of relevant cosmological and particle physics parameters and focusing on their large-scale properties. We have found evidence for scaling behavior of semilocal networks, at least in the sense that both the total string length of the network and the number of string segments evolve towards linear scaling at late times, and we have demonstrated consistency with the predictions of the aforementioned VOS model.

The main bottleneck in previous works, as well as the main source of uncertainties in our modeling---indeed preventing a fully quantitative calibration---is our ability to measure the transverse velocity of the string segments  and the longitudinal velocities of the monopoles in our simulations. Addressing this issue required the introduction and extensive testing of various numerical techniques. The goal of this paper is to describe these techniques, discuss their validation, and present the key results of the analysis of the string and monopole velocities.  It will be seen that our results provide supporting evidence for the scaling of these networks, while also providing additional insights into the physical mechanisms underlying the dynamics of these networks.

 A novel aspect of this work is the introduction of an algorithm that identifies the position of the cores of semilocal string segments much more precisely than in previous works. This allows a more accurate determination of the length and also of the velocity of {\em each}  segment and how these evolve in time, as well as their averages  and distribution across the network. It has revealed some interesting aspects of the velocity distribution, in particular, that had not been reported previously.  We return to this point in the conclusions. In a subsequent paper we will use the knowledge of the evolution of the distribution of lengths and velocities of the semilocal segments and present the results of detailed comparisons of our simulations with the VOS model for semilocal strings, leading to its quantitative calibration.

\section{The model}
\label{model}

The simplest semilocal string model lagrangian reads \cite{Vachaspati:1991dz,Achucarro:1999it}
\be
S=\int\!\! d^4x \sqrt{-g}  \!\left[\left[D\Phi\right]^2-\frac{1}{4}F^2-\lambda(\Phi^+\Phi-\frac{\eta^2}{2})^2\right]\,,
\ee
where $\Phi$ is a doublet of complex scalar fields $\Phi=(\phi_1,\phi_2)$, $D\Phi=(\partial_\mu-i
    q A_\mu)\Phi$,   $F^2 = F_{\mu \nu} F^{\mu \nu}$  and
$F_{\mu\nu} = (\partial_\mu A_\nu - \partial_\nu A_\mu)$ is the gauge
field strength.   The model is invariant under $SU(2)_{\rm global}\times U(1)_{\rm local}$ \cite{Achucarro:1999it}, and after symmetry breaking, the particle content is two Goldstone bosons, one scalar boson with mass $m_s=\sqrt{2\lambda}\eta$ and a vector boson with mass $m_v=q\eta$. After suitable rescalings:
\begin{equation}
\Phi\to\frac{\eta}{\sqrt{2}}\Phi\,,\quad x\to \frac{\sqrt{2}}{q\eta}x\,,\quad A_\mu\to \frac{\eta}{\sqrt{2}}A_\mu\,,
\end{equation}
and defining 
\be
\beta=m_s^2/m_v^2=2\lambda  / q^2\,,
\label{beta}
\ee
the action can be rewritten in the numerically convenient form
\be
S=\int\!\! d^4x \sqrt{-g}  \!\left[\left[D\Phi\right]^2-\frac{1}{4}F^2-\frac{\beta}{2}(\Phi^+\Phi-1)^2\right]\,,
\label{SLaction}
\ee
where now $D\Phi=(\partial_\mu-i A_\mu)\Phi$.

Since our aim is to characterize the dynamics of a  network of semilocal strings in the early universe,  we consider a spatially  flat Friedmann-Robertson-Walker space-time with comoving coordinates:
\be
g_{\mu\nu}=a(\tau)^2\eta_{\mu\nu}\,,
\ee
 where $\eta_{\mu\nu}={\rm diag}(-1,1,1,1)$ is the Minkowski metric,  $\tau$ is conformal time and $a(\tau)$ is the cosmic scale factor. The scale factors we will consider are  those corresponding to a radiation dominated universe ($a\propto \tau$) and to a matter dominated universe  ($a\propto \tau^2$).

 The equations of motion for the semilocal model in  the temporal gauge $(A_0=0)$ read
\begin{eqnarray}
\dot\Pi + 2\frac{\dot a }{a}  \Pi -\mathbf{D}^2\Phi + a^2\beta (|\Phi|^2 -1)\Phi &=& 0, \nonumber\\
\partial^\mu   F_{\mu\nu}  - ia^2(\Phi^{\dagger}D_\nu\Phi - D_\nu\Phi^{\dagger}\Phi) &=& 0,
\label{eom}
\end{eqnarray}
where $\dot{} = d / d\tau$ and $\Pi=\dot\Phi$, together with
\begin{equation}
\partial_i  F_{0i} =  i   a^{2} (\Pi^{\dagger}  \Phi - \Pi^{\dagger} \dot \Phi  ),
\end{equation}
which is Gauss's law.

Semilocal strings can be thought of as Abelian Higgs strings embedded into a larger  model. Indeed, if we set one of the scalar fields in the $\Phi$ doublet to zero, we recover the Abelian Higgs model exactly.  The semilocal model  can also be thought of as a limiting case of the electroweak model where the U(1) symmetry becomes global, and semilocal strings are thus related to the $Z$-strings of the electroweak model \cite{Nambu:1977ag,Vachaspati:1992fi,Urrestilla:2001dd,Achucarro:2005tu}.

Unlike the Abelian Higgs strings, semilocal strings are not topological. 
Therefore, semilocal strings do not necessarily have to be closed or infinite,   but they can form finite string segments: they can have ends. The semilocal string segments are line-like concentrations of magnetic flux, and the string ends can be understood as some sort of global monopoles, with long range interactions \cite{Hindmarsh:1992yy}.
Note that these monopoles are not {\it true} global monopoles; they are, after all, string ends that behave  {\it similarly} to global monopoles. Our analysis provides further supporting evidence for this interpretation.

This combination of string plus monopole  provides   semilocal strings with very rich dynamics: they can shrink and disappear, they can join neighbouring segments to create larger ones, or the ends of a segment can join each other  to form a loop.  
 Besides their more complicated dynamics (compared to Abelian Higgs strings or monopoles), the fact that semilocal strings 
are non-topological has also important consequences for the numerical detection of their spatial location:  there is no topological 
constraint that forces the scalar field to be zero at the core of the string, and also there is no unique winding associated with the 
location of the string. One gauge invariant quantity that we will use to decide where a semilocal string lies is  the concentration 
of magnetic energy.

 The stability of semilocal strings is not trivial. Being non-topological, their stability is not warranted by topology but is instead dependent
on dynamical and energetic considerations. It is controlled by the parameter  $\beta$, c.f. Eq.  (\ref{beta}). Semilocal strings are stable  for $\beta<1$, neutrally stable for $\beta=1$ and unstable for $\beta>1$ \cite{Hindmarsh:1991jq}.

\section{Numerical simulations}
\subsection{Numerical setup}

Our main aim is to characterize the network of semilocal defects. In order to do so, we have obtained  the equations of motion   from  the discretized version of  the Hamiltonian corresponding to the action (\ref{SLaction}), and solved them using  standard techniques (lattice-link variables and a staggered leapfrog method) in  $1024^3$   lattices with periodic boundary conditions, as explained in \cite{Achucarro:2005tu,Nunes:2011sf,Achucarro:2013mga}.  The simulations were performed at the COSMOS Consortium supercomputer and i2Basque academic network computing infrastructure.

Since we have periodic boundary conditions, the simulation time has an upper bound, given by  half the light-crossing time; for times longer than that,
spurious boundary effects kick in.
As  is generally the case with this type of numerical simulation, there is a trade-off between dynamical range, spatial resolution and accuracy. One wishes to simulate the dynamics of the system for as long as possible, and therefore needs boxes with a physical size as large as possible. But one also wishes to simulate the equations as  accurately as possible, which means that the lattice spacing has to be small. Thus, a compromise must be reached. We have chosen to use a lattice spacing of $\Delta x=0.5$ and a time-step of $\Delta \tau=0.2$,  as our optimal balance between accuracy and dynamical range.

 We wish to simulate semilocal string network evolution in the early universe, focusing on the radiation and matter domination eras. 
Defect network simulations in expanding backgrounds present us with another difficulty: the physical size of the defects is fixed throughout the 
simulation, but  the size of the simulation box is growing with time.  Equivalently, viewed in comoving coordinates, the box 
size is constant, but the comoving size of the defects shrinks.  Again, this would point to the necessity to simulate the system in { larger 
lattices, but this is not feasible. Instead, this difficulty is overcome by employing the Press-Ryden-Spergel (PRS) algorithm \cite{Press:1989yh} 
whereby the defect cores are made to `artifcially' grow comovingly during the simulation. This is achieved by promoting the parameters of the 
model into time varying variables \cite{Moore:2001px},  so that the (continuum) equations of motion  (\ref{eom})  get modified to:
\begin{eqnarray}
\dot\Pi + 2\frac{\dot a }{a} \Pi -\mathbf{D}^2\Phi + a^{2s}\beta (|\Phi|^2 -1)\Phi &=& 0, \nonumber\\
\partial^\mu \left(a^{2(1-s)}  F_{\mu\nu} \right) - ia^2(\Phi^{\dagger}D_\nu\Phi - D_\nu\Phi^{\dagger}\Phi) &=& 0,
\label{eom2}
\end{eqnarray}
where the parameter $s$ gives the level of modelling  in the PRS algorithm: $s=1$  gives the true equation of motion, whereas $s=0$  corresponds to defects whose  comoving core size is constant. Several studies in the literature \cite{Moore:2001px,Bevis:2006mj,Daverio:2015nva,Lopez-Eiguren:2016jsy} show that the $s=0$ case is a good approximation to the true $s=1$ situation, with errors often smaller than statistical  uncertainties. In the present work we performed all simulations using $s=0$.

We are interested in estimating the velocity of the network, and more precisely the velocity of the segments and of the string--ends (monopoles). But for the purposes of calibrating the VOS model, it is particularly useful to do this  
once the system  has reached scaling. As mentioned in the Introduction, scaling is a convenient
property for the reliable study of defect networks,  since we need to extrapolate from our limited numerical simulation into the much larger timescales relevant for defect networks evolving in the real universe. For scaling networks, the details of the initial conditions 
are not important as all information about the initial condition is lost once scaling has been reached. However, choosing a `good' set of initial conditions is important in practice, as it can help the network reach scaling as fast as possible, so that the system can evolve in the scaling regime for 
as long as possible within our finite dynamical range. The initial conditions we chose are the same as those in \cite{Achucarro:2013mga}: 
the gauge field, gauge field velocities and scalar field velocities are set to zero. The scalar fields are chosen to lie in the vacuum manifold, but 
have randomly chosen orientations \cite{Vachaspati:1984dz}. Note that one extra benefit of using the PRS algorithm with $s=0$ is that scaling is reached faster, which 
 further increases our usable dynamical range.

We want to simulate strings that are stable and lead to long enough segments, so that we can get a representative statistical ensemble in our simulations. As mentioned earlier, semilocal strings are stable for  values of $\beta<1$, and our simulations were performed for $\beta=0.04,0.09,0.15,0.20,0.25,0.30,0.35$. Higher values of $\beta$ show a much more scarce network \cite{Achucarro:2005tu}, and it would be much harder to get a  numerous enough 
ensemble of semilocal strings. Lower values of $\beta$ are also not optimal, since the scalar string cores start to be too  thick for our simulation parameters.  

For every value of $\beta$, and for both cosmologies (radiation and matter), we  have performed 7 different simulations, in order to increase our statistics. It is   numerically very expensive to analyse and output the  simulation data at every time-step; instead,   the information is output every 20 time steps once the network has approached scaling. The times  chosen to output the data range from $\tau= 96$ until $\tau=256$;  for times earlier than $\tau=96$  the system has not settled into scaling well enough, and $\tau=256$ is our limit due to the aforementioned boundary conditions. For a couple of cases the information has been output more  frequently too, in order to check that the time spacing between outputs was adequate, and also to pinpoint some issues that we will discuss later. The  output information can be used directly to estimate some of the  quantities of interest, and also can be treated further to infer other  useful quantities, as explained in the next section.

\subsection{Algorithms to detect segments and obtain their velocities}

In this subsection we will describe the quantities that need to be monitored in order to characterise semilocal string networks. We will introduce algorithms for estimating string length, string velocities and monopole velocities; some of these algorithms have already been used previously in the literature and some are introduced here for the first time and are specific to semilocal strings. Note that this is a general description; various caveats will be described in more detail in the following section.  

As mentioned   earlier, the fact that semilocal strings are not topological makes it much more difficult to numerically detect their location 
in a simulated lattice compared to the (much more studied)  AH strings.  In the AH case, the strings have a topological constraint, and one can follow 
 them  throughout the simulated lattice by tracking the winding of the scalar field across each plaquette \cite{Kajantie:1998bg}. The points with winding coincide with the zeros of the scalar field
  and  therefore the windings happen at the same points  at which the potential energy is concentrated. Furthermore, those 
are   also the regions where the density of magnetic energy is    highest. Thus, one can use any one of these features 
(high magnetic field density, high potential energy density, zeros of the scalar field, winding of the scalar field)  to detect an AH string.

In the semilocal case, there are twice as many scalar fields, and just one gauge field. It is not clear then which field one has to choose to follow 
the windings, or whether these windings actually give  the position of the string. Besides, the field does not have to  be zero when there 
is a   winding. Unlike in the AH case,  {\it one}  field can be winding around a site, but the {\it other} field may still fulfil the requirements 
to be in the vacuum of the potential energy. Therefore, we can have a field winding, but no concentration of potential or magnetic field.

In previous works \cite{Achucarro:1998ux,Achucarro:2005tu,Nunes:2011sf,Achucarro:2013mga},  the criterion used to decide whether a point 
in the lattice  belongs  to a semilocal string was based on the concentration of magnetic   energy (we will describe this 
method in more detail below).  The procedure gives 
a collection of points that  can be grouped  into segments by proximity, {\it i.e.} we end up with a {\it volume} of    points, which is 
subsequently used to  estimate length.  This length can be used to check, for example, the scaling of the network of strings. Moreover, the number 
of segments is also a good estimator for the number of monopoles, {\it i.e.} segment ends, which is a further diagnostic for the scaling of the network. It has always been understood  that this was just a first approach to obtain the length of strings, and we will show later that this estimator was seriously underestimating the string length.

Instead of a volume characterization of the strings, one would rather get a one-dimensional    representation of strings, using, for example, 
the position  of the core of the string. We will introduce a new estimator using field windings that will produce    such a one-dimensional 
characterization of the string.  This will be invaluable for the estimators of velocity we will describe    later, and will enable us to determine 
individual segment velocities for semilocal strings  for the first time.

We have also detected monopoles directly from the simulation for the first time. The number and velocity   of monopoles 
are also important quantities for characterising our system. For example, from the total number of monopoles $\mathcal{N}$  an independent 
measurement of the number of segments can be inferred.

In what follows we will describe the algorithms used to obtain lengths of segments, number of segments, number of monopoles; 
velocities of   segments and monopoles, network velocity; and quantities used  to monitor scaling.

\begin{itemize}
\item Estimation of segment lengths using the threshold of the magnetic field:

One criterion  for deciding  whether a lattice-point in the simulation   is part of a semilocal string is the following: 
points where the magnetic field is higher than a given threshold are considered to belong to a semilocal    string. In order to get a meaningful threshold, for a given $\beta$ we consider the corresponding Abelian Higgs string, and calculate the maximum of its magnetic field $B_{\rm max}$. We also calculate the radius of the Abelian Higgs string  $r_{\rm AH}$, defined as the radius at which   the magnetic  field $B_{threshold}$ (absolute value) drops to 30\% of $B_{\rm max}$. 
Armed with the value $B_{\rm max}$,  the whole simulation grid is   scanned at every time--step,  the value of the magnetic field is calculated at each grid point, and if the magnetic field is higher than 
that of $B_{threshold}$, the position of that point is   output.

Once all the points have been output, they  are distributed into segments. In order to do that, the points that are adjacent to each other 
are grouped into one segment. Thus, a collection of points forming the volume of a segment is obtained. Dividing that volume 
by the  cross-section area of the string (calculated from $r_{\rm AH}$ assuming a 
circular cross-section), the length of the segment is estimated. From this procedure a distribution of segment lengths is obtained at every 
time step. Adding up all the segments, the total length of string in every time step $\mathcal{L}(\tau)$  can also be obtained, which can be 
used, for example, as a measure of scaling.

The left  panel of Fig.~\ref{fig:example} shows the positions of the segments in a typical simulation  box obtained by plotting 
the points with a magnetic field higher than 30\% of $B_{\rm max}$. This was the procedure used in previous papers to obtain semilocal segment lengths \cite{Achucarro:2005tu,Nunes:2011sf,Achucarro:2013mga}.

\item Estimation of segment lengths using the windings of the scalar fields:

This is a new length estimator which allows us  to obtain a one-dimensional characterization of the segments. In order to do so, it uses the  windings in either  
  or both of the scalar fields ($\phi_1$ and $\phi_2$) in  the model.  

The procedure is as   follows:  During the simulation, the winding of fields $\phi_1$ and $\phi_2$ is calculated at every point, and if this winding   is  different from zero, that  position    is  output   together with the value of the magnetic field at that point. The value of the magnetic field is needed because  having the fields  winding around some plaquette  does not necessarily mean that point belongs to a string segment; it is only when the non-zero winding happens in a region with a high concentration of magnetic field that the position is regarded as the core of a string. The position of the core is given by the non-zero winding of either (or both) of the scalar fields (within the cloud of points with magnetic field), and following the points with winding one   can map the 1-D array of points defining the centre of the string. 

Actually, since there is no topological constraint and  we are following the points with non-zero windings of one of the fields, the points may stop belonging to a semilocal string segment (because there may not be a high concentration of magnetic field), or   the sequence of points with windings may abruptly stop. It is worth noting that, in principle, for a point to be part of  a semilocal string it is enough that one of the fields winds in a region with high magnetic field. It is not necessary that the other field also winds. (Nevertheless, we will show later on that both fields seem to wind inside the string core.) On the other hand, sometimes both fields may be winding inside a cloud of magnetic field, but  (due to discretization issues) not exactly on the same plaquette   and they  may be displaced by one lattice unit.  In order to overcome these nuisances, the points of non-zero windings of both fields are combined.  Those points that  share neighbours are  grouped, and that is how we define a segment. Moreover, those points can now be ordered . Every segment is thus formed of a (one-dimensional) ordered collection of points where either (or both) of the scalar fields have a  non-zero winding and magnetic field is above some threshold,  and points are in contact with each other by pairs.   Figure~\ref{fig:example}   shows the string   positions, obtained by taking  into account the concentration of magnetic field (blue) and the points where the fields are winding ($\phi_1$ in red, $\phi_2$ in cyan and both in black).

Since our fields are discretized on a lattice,   the determination of the centre  of the string suffers from the   Manhattan effect \cite{Scherrer:1997sq}: the center of the segment is not a smooth curve, but it is actually formed  by a collection of unit steps.  A smooth version of the position of the cores would be   beneficial as the estimation of both the length and the velocity of the segment would improve.  Thus, 
the points in the string segments are smoothed  by averaging the position over nearest neighbours. After   trials using different number of neighbours, averaging over 4 neighbours at each side seems optimal: the position of the core is   smooth  enough, and  the  structure of the segment is not lost (as would happen by averaging over  too long distances). 

The outcome of this procedure is a collection of segments, given by a one-dimensional list of  the smoothed position of their core.  The right   panel  of Fig.~\ref{fig:example} depicts a zoom into a portion of a segment, showing how the averaged version of the core of the string, obtained by smoothing,  nicely interpolates between the points with non-zero windings.

\begin{figure}[!htbp]
   \centering
      \includegraphics[width=16cm]{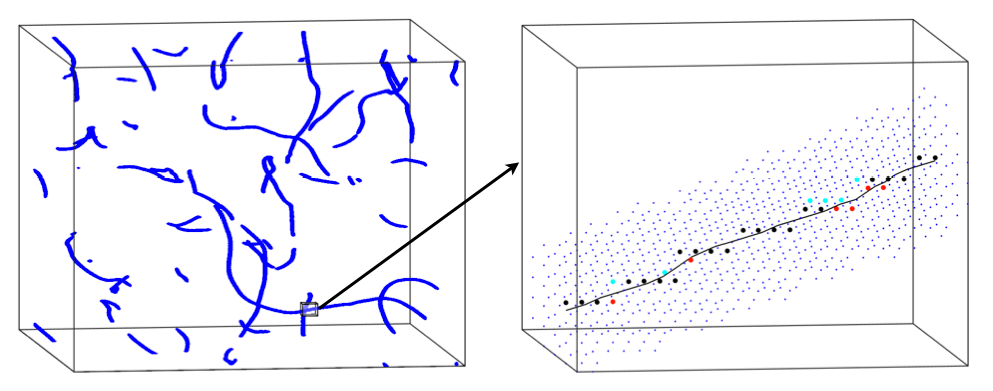}
   \caption{   \label{fig:example} In the left   panel, the string segments in a typical simulation box are shown;  the blue points are points with a magnetic field higher than the threshold described in the text. The right   panel  zooms in on a portion of one segment. The blue dots correspond again to points with high magnetic field, and the  red, cyan and and black   points correspond to points where the windings of the scalar fields $\phi_1$, $\phi_2$  or both (respectively)   are non-trivial.    Notice how   these  dots are not nicely aligned in a smooth curve, but jitter around (this is the  Manhattan  effect). Performing the smoothing over four nearest neighbours on each side of each point yields the black line, which is the smoothed version of the  centre  of the core of the segment.  Actually, the black line is also formed of points, but here  those points  have been joined by a line to show how they interpolate between the lattice--points with winding.} 
\end{figure}

\item Estimation of the velocity of the segments:

Once the  segments have been characterized by a smooth one-dimensional array of   positions, this set of data  can be used to  track the history of each segment in time, and also estimate   its (transverse) string  velocity.  

Consider  a string segment at time $\tau_1$, and the same segment at a later time $\tau_2$. If one is able to estimate where each point in the segment has moved from $\tau_1$ to $\tau_2$, we can get an ensemble average. This can be done in the following way: choose a point $x_1$ in the core of a segment at   time  $\tau_1$, and  find  its distance to all points belonging to  the segment's core  at  $\tau_2$. The point at   time  $\tau_2$ that is closest to $x_1$  is identified  as the point where   $x_1$ has moved to from $\tau_1$ to $\tau_2$. \footnote{We do not remove points at the later stage, so that it may happen (and it will clearly happen for example in a collapsing loop) that two or more points at $\tau_1$ may move to the same point $\tau_2$.} The velocity of point $x_1$ is then estimated by merely dividing  the distance it has travelled by  the time interval $\tau_2-\tau_1$. Performing this operation with every point belonging to a given segment at every time step allows us to obtain a segment velocity (an average of $|v|$ over all the points in the segment). The average of all the segment velocities, which is not length-weighted, would then be the network average velocity.

Although this procedure has some   subtleties, which we will describe in the next section, the benefits are manifold:  one can obtain individual segment velocities at every time  step. We can thus record the history of the velocity evolution of each segment in the simulation.  Moreover, a  by-product of this procedure is that  one can create a map of which segments merge with which.

It is clear that for this procedure to work a one-dimensional characterization of the string segment is necessary, and therefore  the velocity determination of segments relies  on the string length characterization given by the windings.

\item Estimation of number and velocity of string-ends (monopoles)

As mentioned in section~\ref{model}, the field configurations at the ends of the segments  can be understood as global monopoles \cite{Achucarro:1999it} and, as such, they  can be detected in a simulation directly from the fields. The whole simulation lattice is  scanned point by point to check whether monopoles are present. In order to do so,  following \cite{Achucarro:1999it}, the field configurations at the string ends are recast from the 4 real scalar fields with SU(2) symmetry into three scalar fields,  establishing an analogy to  $O(3)$ global monopoles,   through 
\be
\Psi \sim \Phi^{\dagger} \vec{\sigma}\Phi,
\ee
where $\vec{\sigma}$ are the Pauli matrices.  Actually, the field configuration thus obtained is still quite   noisy  to clearly detect monopoles, and we clean it further via the transformation
$$  \Psi\ \longrightarrow \ \ \tilde{\Psi}=\Psi-\nabla \times (\nabla \times \Psi)\,.$$
Once the field configuration has been treated, the location of a monopole is determined by calculating the  topological charges given by the surface integral \cite{Vilenkin:2000jqa}
\be
N=\frac{1}{8\pi}\oint dS^{ij}|\tilde{\Psi}|^{-3}\epsilon_{abc}\tilde{\Psi}^a\partial_i\tilde{\Psi}^b\partial_j\tilde{\Psi}^c.
\label{tp}
\ee
where  $dS^{ij}$ is the infinitesimal surface element.  Since we are in a discretized environment, Eq. (\ref{tp}) cannot be used directly, and instead a discretized version  is used (see Appendix B  in \cite{Antunes:2002ss}). Note that $N$ can be positive (monopole) or negative (antimonopole). It is not to be confused with $\mathcal{N}$, the total number of monopoles plus antimonopoles, i.e. the total number of segments' ends.

When scanning through the simulation box, if the topological charge is different from  zero, the position of the monopole together with its topological charge is output.  One can then directly obtain the number of monopoles (or string ends) in the simulation lattice, which is an independent way of measuring the number of string segments: the number of segments would be roughly  $\mathcal{N}/2$, although not exactly so because some segments can be closed loops  and thus have no ends. Moreover, the number of monopoles is another estimator   for checking  the scaling regime.

Since   the position of the monopoles (and their topological charge) is known at every time-step, the velocity of monopoles  can also be estimated following a  procedure similar to the one described above for segment velocities \cite{Lopez-Eiguren:2016jsy}: A monopole $M_1$ is chosen at  time $\tau_1$, and its distance with respect to   all monopoles  at the next time $\tau_2$ is calculated. Then, $M_1$ is identified with the closest monopole at time $\tau_2$. 
Repeating this procedure at all times,  the history of $M_1$ can be   tracked  and  an estimate of its  velocity can be obtained by merely dividing the distance   it has travelled by the corresponding time interval.

 Monopole velocity estimations  also have their issues, but once again the benefits are manifold: one can obtain individual monopole velocities, both at   each time step and as the average velocity during the life  of the monopole. By averaging over all monopoles, a monopole-network velocity can be obtained too.

\item Estimation of the network velocity using local lattice variables.

There are also other   types of velocity estimators  for the whole network, based on local field quantities   as described in  \cite{Hindmarsh:2008dw,Hindmarsh:2016dha,Hindmarsh:2017qff}.   These estimators are computed by considering a string at rest, and performing a Lorentz boost to it. Thanks to the different properties of each term in the Lagrangian under boosts,  two different estimators can be written down:
  \begin{eqnarray}
&&\langle v^2 \rangle _F=\frac{\mathbf{E}^2_{\mathcal{W}}}{\mathbf{B}^2_{\mathcal{W}}},\nonumber\\
&&\langle v^2 \rangle_G=\frac{2G_{\mathcal{W}}}{1+G_{\mathcal{W}}},
\label{Markestimators}
\end{eqnarray}
where $F$ refers to the gauge field strength $F_{\mu\nu}$, and
\be
G_{\mathcal{W}}=\frac{ \Pi^2_{\mathcal{W}} }{(\mathbf{D}\Phi)^2_\mathcal{W}}.
\ee
 
The subscript $\mathcal{W}$ denotes weighting by some   appropriate physical quantity. In the present case, a magnetic energy weighting was used, because   this automatically ensures that only regions with non-vanishing magnetic energy contribute to velocities.   Regions with semilocal strings have higher concentration of magnetic energy, and thus those regions contribute most  to the above integrals. For a given quantity $A$ the weighting is applied in the following way:
\be
A_{\mathcal{W}}=\frac{\int d^3 x A \mathcal{W}}{\int d^3 x \mathcal{W}}
\ee 
Using this definition for a {\it weighted average}, the estimations above read  
$$G_{\mathcal{W}}=\frac{ \Pi^2_{\mathcal{W}} }{(\mathbf{D}\Phi)^2_\mathcal{W}}=\frac{\int d^3 x \Pi^2 \mathcal{W}}{\int d^3 x (\mathbf{D}\Phi)^2 \mathcal{W}}$$
for example. During the simulation, the quantities  needed to obtain the velocity estimators in (\ref{Markestimators}) are output.

Note that these field velocity estimators provide information on the network as a whole, whereas the previously described methods for segments (and monopoles) give information about each individual segment (or monopole).
Besides, the estimators in Eq. (\ref{Markestimators}) average over all regions with some magnetic energy, and thus both segments and monopoles  contribute, as well as  regions where a segment may have disappeared (or it is about to disappear) and a temporary magnetic field density is left-over.

\item Monitoring scaling:

Scaling is a key property of a network of defects, which is indispensable if one needs to extrapolate the results of a numerical simulation (whose dynamic range will necessarily be small for the scales of interest) into the history of a realistic defect network.
Scaling can be measured by the total string length $\mathcal{L}(\tau)$ and by the number of monopoles $\mathcal{N}(\tau)$ (which is related to the number of segments, $\approx \mathcal {N} (\tau) / 2$). Besides, the energy ($T_{00}$) of the system is also a good candidate to monitor the  scaling of the  network.
  
\end{itemize}

\section{Caveats and difficulties of the algorithms and the numerical setup}

In this section we will  describe some caveats and difficulties we encountered in the simulations. These, on the one hand, have to be dealt with in order to obtain physically robust results, and on the other, provide insights into how to further extend the analytic model for semilocal strings (the last point will be addressed in a subsequent paper). 

\subsection{Comparison between the two length estimators}

In the previous sections,  two different procedures   for obtaining the length of string segments  have been described: one using the threshold of the magnetic field  and   the other   using the windings of the scalar fields. 
We did expect some mismatch between   these two procedures,  mostly because the method of obtaining segment lengths by counting points above a magnetic field   threshold and dividing the {\it volume} by the {\it cross-section area} was a rather  crude algorithm. This algorithm was used in previous works  \cite{Achucarro:2005tu,Nunes:2011sf,Achucarro:2013mga}  due to the difficulty of defining a semilocal string otherwise (with windings or zeros of the scalar field), and was believed to be a good first approximation.

\begin{figure}[!htbp]
   \centering
   \includegraphics[width=8cm]{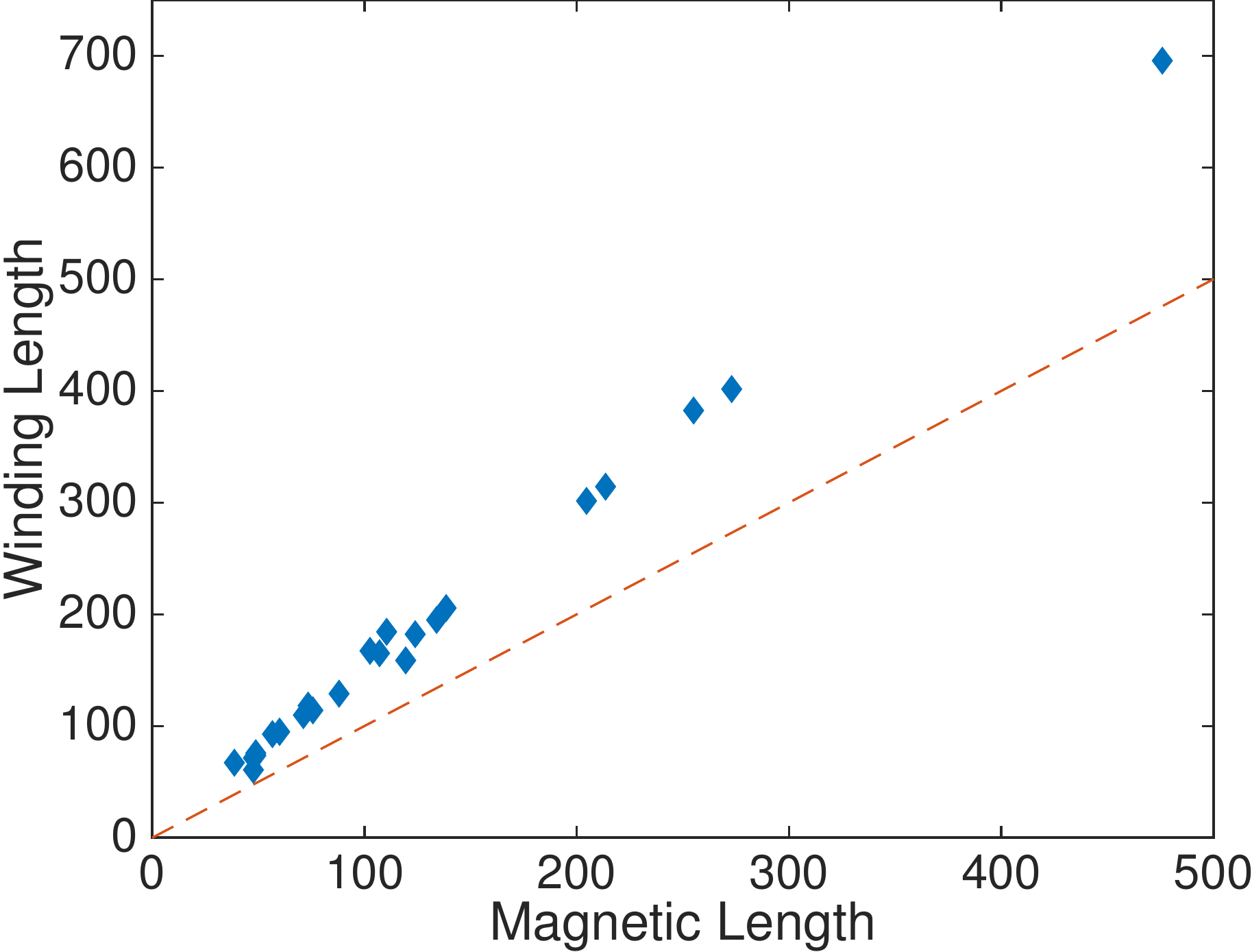} 
 \caption{\label{seg-rel}In this figure the relation between the two length-measuring approaches is shown. The horizontal axis has the lengths measured with the magnetic flux approach, while the vertical axis has the lengths measured with the winding approach. Every diamond represents a segment in a simulation box for $\beta=0.20$  at time $\tau=256$.  It is clear that the proportionality is almost the same for all the segments in the network. The diagonal is shown with a dashed line, emphasizing the bias factor between the two approaches.}
\end{figure}

Fig~\ref{seg-rel} shows the relation between the lengths of the segments measured by using the magnetic flux and the winding approach for $\beta=0.20$ at $\tau=256$.  The ratio between these two quantities is almost constant, and it is roughly a   (surprisingly large) factor of 1.5 for most values of $\beta$ both in the matter and radiation cases.  The factor is a bit different for shorter  segments, which is  presumably due to the fact that the segment ends play a stronger role for shorter segments, but this is not relevant for the whole network   length, which is dominated by long segments. 

\begin{table*}
\begin{center}
\begin{tabular}{|c|c|c||c|c|}
\hline
 & \multicolumn{2}{c||}{Radiation} & \multicolumn{2}{c|}{Matter} \\
 \hline
 $\beta$ & $L_{winding}$ & $L_{magnetic}$ &  $L_{winding}$ & $L_{magnetic}$ \\  \hline

 0.04 & 6166.44  & 3895.33  & 12303.07  & 8014.28   \\  \hline
 
 0.09 & 6903.84  & 4673.61  & 10780.82  & 7266.59   \\  \hline
 
 0.15 & 5246.97  & 3476.99  & 9267.51  & 6099.45   \\  \hline
 
 0.20 & 5221.27  & 3546.95  & 7447.93  & 4865.46   \\  \hline
 
 0.25 & 4450.95  & 2948.52  & 6147.42  & 3853.46   \\  \hline
 
 0.30 & 2546.31  & 1585.21  & 4562.78  & 2901.85   \\  \hline
 
 0.35 & 2095.5    & 1173.86  & 3405.10  & 2119.90   \\  \hline

\end{tabular}
\caption{\label{tot-length} The total lengths in $\tau=256$ for every $\beta$ we simulated and for Radiation and Matter epochs. $L_{winding}$ denotes the length estimator using the windings of the scalar fields, whereas $L_{magnetic}$ denotes the length estimator using a threshold of magnetic field. Note that these numbers correspond to a single simulation per value of $\beta$.}
\end{center}
\end{table*}

There are several possible sources of uncertainty that can help us understand the difference in the string length obtained with the different procedures. First of all, previous determinations discarded small ``blobs", very short segments whose length is below some threshold (a few times the core radius).  Second, the fact  that cross-sections are estimated from the Abelian Higgs profile could lead to an overall bias. There are also two possible physical effects to take into account: one is Lorentz contraction of the segments due to their velocities, making segments narrower than the estimated   cross-section  of the string; the other is that segments are not straight, and when they bend the strings can be considerably  narrower. 

There is also a clear numerical uncertainty in the procedure of obtaining the cross-section  of the string. Recall that the cross-sections of the semilocal strings in the simulations are defined to be the same as those  of a straight AH string at rest. The typical radius  for AH strings range from $r_{04}=2.36$ for $\beta=0.04$ to $r_{35}=1.71$ for $\beta=0.35$. The lattice spacing we are using is $\Delta x=0.5$  which already gives   us a considerable error on the string radius: we consider that points with magnetic field higher than 30\% of the maximum magnetic field belong to the string. Imagine a string centered at a lattice-point, with radius, say,  1.4. The string radius then does not quite cover 3 lattice units ($3*\Delta x>1.5$), and therefore there are only 2 lattice points which {\it qualify} as points of strings, but the volume of strings obtained like that will nevertheless be divided by the {\it true} radius, clearly underestimating the string length.  Moreover, a string will typically not be centered around a lattice--point, but it will be located anywhere within a lattice--cell, making this effect more important. Thus, the points that we {\it numerically} decide that belong to the string are {\it fewer} than what a more sophisticated procedure would get. We can get an estimation of the maximum errors for the two extreme $\beta$ as follows:
$$\frac{r_{04}^2}{(r_{04}-\Delta x)^2}=1.60\,,\qquad \frac{r_{35}^2}{(r_{35}-\Delta x)^2}=1.99\,.$$

This numerical bias seems to account for most of the discrepancy between the two length estimation procedures, while the Lorentz factor and  the fact that strings are not straight do not seem to be so important. Actually, the Lorentz contraction of the segments is clearly not a significant factor to take into account since the velocity values we obtain later on are rather low to obtain a Lorentz factor capable of explaining the difference. In Table~\ref{tot-length}, the total length in segments for every $\beta$ is shown. We conclude then that the string segments length estimated in previous works by (some of) us  has been overestimated by roughly a factor of 1.5 (the average ratio between the values in Table~\ref{tot-length} are 1.56  for Radiation and 1.55  for matter).

When using the approach based on obtaining volumes of points and then dividing by string cross-sections, there was also some nuisance with very small segments, which were dubbed {\it blobs} in \cite{Achucarro:2013mga}. These were very small regions with some magnetic flux, possibly as a leftover of a recently collapsed segment. There was some uncertainty in how to remove those blobs. With the new winding approach, this problem of blobs has disappeared. In any case, we have checked that the number of monopoles obtained by the winding approach, and by the magnetic field approach (removing blobs)  is very similar.

One very nice check that our new length estimators work well can be found in Figure~\ref{together}. There, we plot the positions of the strings obtained by the winding procedure, and the positions of the monopoles obtained by calculating the monopole charge. In principle, these two procedures are independent, but  the positions of the strings and monopoles (string--ends) are aligned very nicely.

\begin{figure}[!htbp]
   \centering
   \includegraphics[width=10cm]{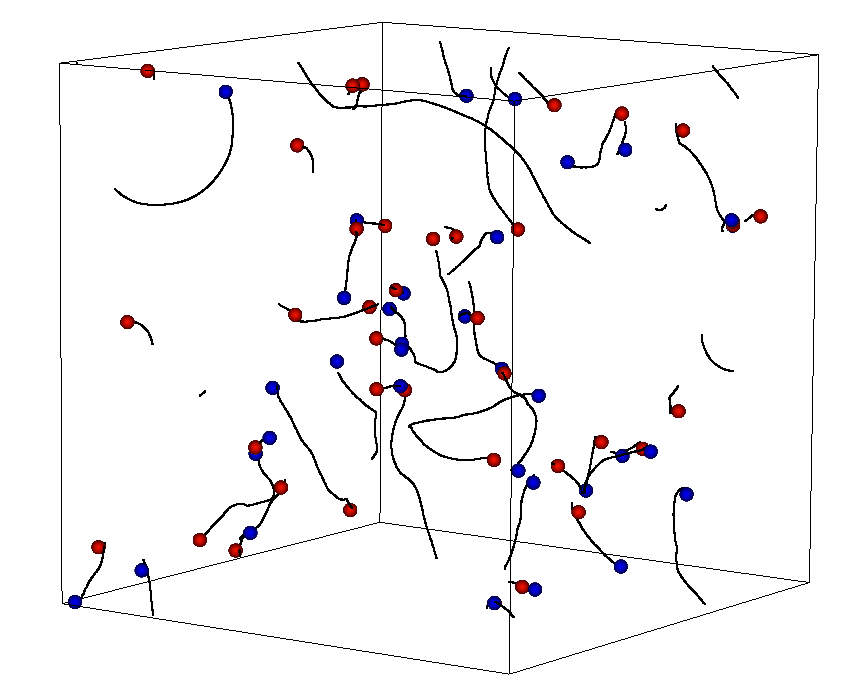} 
     \caption{ \label{together} In this figure one can see the box after the characterization. The black lines represent the string segments detected using the windings, where each point is smoothed using the four nearest neighbours in each direction. The red dots represent the antimonopoles and the blue ones the monopoles.  }
\end{figure}

From the length estimators (of both monopoles and strings)  two different VOS-type length scale parameters can be obtained:
\begin{equation}\gamma_{\mathcal{L}}\equiv\frac{1}{\tau}\sqrt{\frac{V}{\mathcal{L}}}\,,\qquad \qquad\gamma_\mathcal{M}\equiv\frac{1}{\tau}\left(\frac{V}{\mathcal{N}}\right)^{1/3}\label{params}
\end{equation}
where $\mathcal{L}$ is the estimation of the string length, and $\mathcal{N}$ is the estimation of number of monopoles (double the number of segments).  We show the values of those parameters in Table~\ref{tab-gamma} in Section~\ref{results}, but we would like to point out that 
these values are compatible (within errors) with the values obtained in \cite{Achucarro:2013mga}, for both $\gamma_{\mathcal{L}}$ and $\gamma_{\mathcal{M}}$. 
Note that, in view of our results, the values for $\gamma_{\mathcal{L}}$ in  \cite{Achucarro:2013mga}  should be corrected by a $\frac{1}{\sqrt{1.5}}$ coming from the  factor $1.5$ difference in the length estimator,  but even without the factor, the values lie within 1-$\sigma$ from each other. 
 
In the rest of this work, we will only use  the length estimator  which relies on the  windings approach. The reason   for calculating  segment lengths using the threshold of the magnetic field was for ease of comparing with previous works; it will not be used in the rest of this paper.

\subsection{Identification of semilocal segments}

As has been already explained,  the position of a semilocal string is difficult to pinpoint because there is no topological obstruction 
for the scalar to acquire a non-zero value around a winding (one field may wind and the other may climb the  potential) and there is no requirement 
for semilocal strings to be closed or infinite. Actually, it is  {\it finite semilocal string segments} that we are studying.

One way of detecting the  strings, explained above, is by following the windings of  both scalar fields $\phi_1$ and $\phi_2$, and checking where those windings   happen within regions of high magnetic field. In the left  panel of Fig~(\ref{fig-pq})  we show the windings of both fields in the simulation box, without taking into account the values of the magnetic field: windings of $\phi_1$ in blue and of $\phi_2$ in green. Actually, there are several regions where the two fields wind in the same plaquette (or at a difference of 1 plaquette), and those are 
plotted in red.

\begin{figure}[!htbp]
   \centering
   \includegraphics[width=8cm]{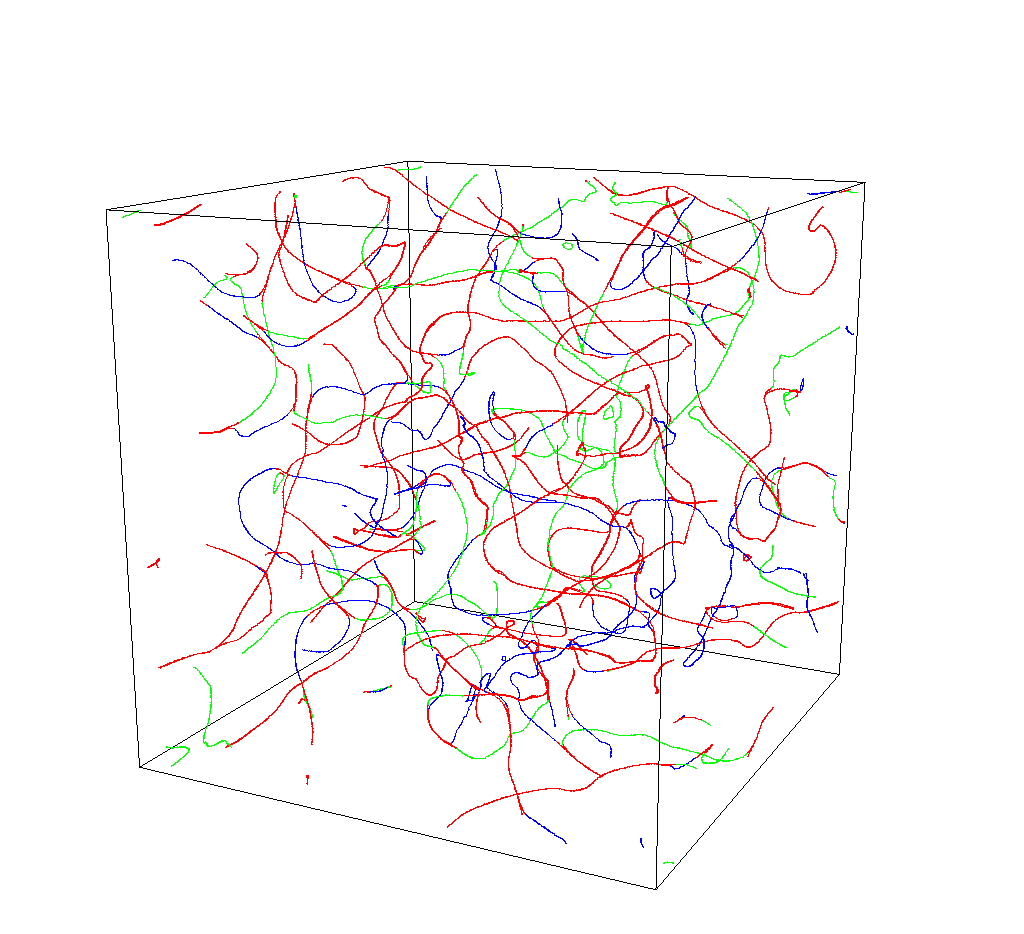} 
   \includegraphics[width=8cm]{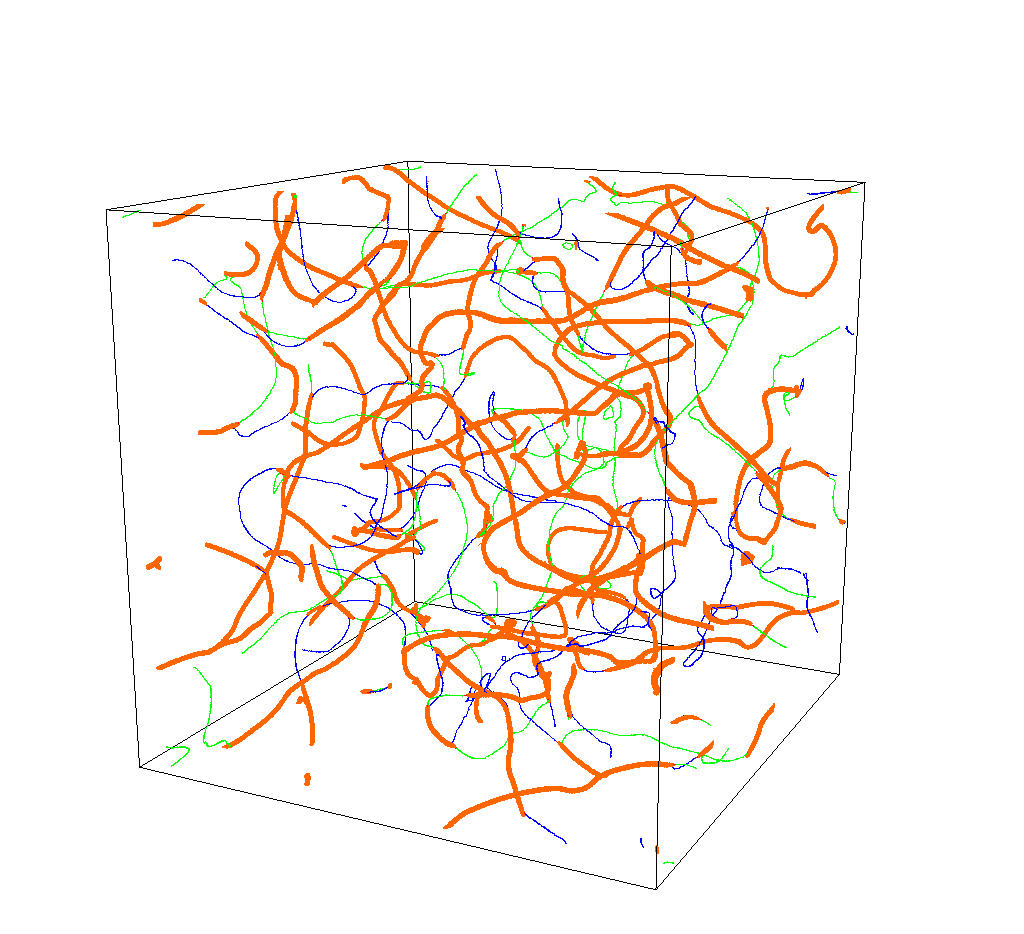} 
     \caption{ \label{fig-pq} In the left panel the windings of both fields are shown, without taking into account the magnetic field: blue for windings of $\phi_1$, green for windings in $\phi_2$ and red for windings in both. As one can see each one of the fields creates a network of closed or infinite strings. On the right  panel, we plot the same windings, but also the regions where there is concentration of magnetic field  (orange). This figure shows that semilocal strings are actually the regions where  both fields  wind. }
\end{figure}

The first thing we notice is that each  of the fields ($\phi_1$ and $\phi_2$) creates a network of closed or infinite  `strings', very much like an Abelian Higgs network.  But unlike in the Abelian Higgs case, a winding of a scalar field does not necessarily mean that that there is a string forming; there is not necessarily a corresponding concentration of magnetic or potential energy. Where do semilocal strings form then?
 In the right  panel we plot the same windings, but also the regions where there is   a high concentration of magnetic field. This shows that semilocal strings are actually the regions where {\it both} fields wind,  which makes sense because in that case the scalar field has to be approximately zero there and therefore the magnetic field can be large in those regions.
 
Note that, although this network  formed from points with windings of $\phi_1$, points with windings of $\phi_2$  and semilocal strings is reminiscent of a p-q string network \cite{Copeland:2003bj,Dvali:2003zj}, where  the Y-junctions correspond to monopoles, this analogy is misleading.  The winding of a single field does not necessarily imply any physical concentration of energy, so there is no string tension that can be associated unambiguously with these ``strings".

\subsection{Apparent superluminal velocities}\label{superluminal}

The preliminary results on velocities (for both segments and monopoles) obtained by most of the direct algorithms, as well as from estimators based on local field quantities, cf. Eq. (\ref{Markestimators}) give rather modest velocities. However,   on  a few occasions, we obtained values higher than the speed of light c; in some very few extreme cases as large as $10c$.  In principle, as we explain below, this need not signal a breakdown of causality, and it can be an artefact of the way we measure velocities. The dynamics is local and causal at the level of the fields.

A more detailed analysis showed that superluminal values for velocities  were only obtained when using  the estimators (described above) which try to track the segments and the monopoles in each time step as the network evolves. We also found out that the procedure needed some more careful handling, due to the following caveats:

\begin{itemize}
\item Segment velocities:

Imagine a segment that has closed into a loop at time $\tau_1$, and disappears before the next  time  $\tau_2$ when the box will be re-analysed. Then, all the points belonging to the loop at time $\tau_1$ will not have a close-by point to match with at $\tau_2$. However, our code will find the closest segment at time $\tau_2$, even though that will correspond to another segment, presumably further than causality allows. Therefore, the calculated velocity for the loop will be very high, but clearly   it is not physical; it is a shortcoming of the procedure. One way of dealing with this is by applying a cut-off, and disregarding all velocities higher than that cut-off, but this has the danger of possibly losing some dynamics (if the cutoff is too strict), or considering too many  unphysical cases (if the cutoff is too weak). Another possibility to avoid these problems is by performing the identification  between segments at different times backwards, i.e., choosing a point at $\tau_2$ and looking at all points at $\tau_1$ to match it to.

However, if we revert to the {\it backward} identification, we may run into another problem: Imagine now a segment that is going to join to another segment, and for the sake of explaining   this issue assume that, with the exception of the segment's ends (which are traveling towards each other), the string is at rest.
At time $\tau_1$ the two segments are separated by some distance, but at time $\tau_2$ all that distance is filled with string. All the points that have filled that gap at time $\tau_2$ will have their closest point at the ends of the segments  at time $\tau_1$ and therefore the velocity we get for those points is not the velocity of the segment. If at all, that velocity will be related to the segment end (monopole) velocity. This is avoided by calculating  velocities {\it forward} (from $\tau_1$ to $\tau_2$). 

Clearly, choosing the forward or backward approach may  solve one of the two previous problems, but not both.

\item Monopole velocities:

Throughout the simulation monopole-antimonopole pairs annihilate: if a monopole $M_1$ annihilates with an antimonopole   between times $\tau_1$   and $\tau_2$, our procedure will not have the information to know that monopole $M_1$ has annihilated. Instead, the procedure will try to match it to another monopole (the one nearest to it), and the velocity obtained will be incorrect (and possibly supeluminal).

Other types of mis-identification issues may also occur occasionally:  There are some detection problems when a string segment collapses into itself, as the fields are  reconfiguring to the new situation and radiating energy away. In this case, the field configuration at the string ends is related to the merging of a monopole and an antimonopole, and our detection algorithm fails (either by detecting spurious monopoles, or by failing to detect any charge at all).  

Moreover, if at the moment when the identification algorithm is run a monopole passes through a face of the lattice, the algorithm may miss to detect it, because the topological charge will be divided into two different cells of the lattice.   The way  we have chosen to overcome this difficulty is by checking whether the monopole can be matched to a monopole in a subsequent time-step $\tau_3$, and to compare the distances obtained. If the distance to the monopole at $\tau_3$ is smaller than to that at $\tau_2$ (and is actually physical), the intermediate step is by-passed.  The fact that the charge of the monopole is recorded is of help when there are monopoles and antimonopoles nearby: the topological charge of a monopole does not change during its  evolution, {\it i.e.} a monopole does not become an antimonopole, and therefore the number of possible candidates to match to a given monopole is halved.

\end{itemize}

After this refinement of segment and monopole velocity estimators, most of the instances where superluminal velocities were obtained   have been cured. However,  there remains some cases where superluminal velocities appear, corresponding to a few instances of merging of segments as mentioned above. These superluminal velocities appear at a few points of a string segment close to the end of the string, and also obviously at the monopoles. For the segments of strings, the superluminal velocities appeared only in the backward direction, and their effect is to increase artificially the velocity of the segment (since the segment velocity is an average of the velocities of the segment points). 

We individually looked at those cases, in order to determine whether those high velocities where physical or had their origin on the shortcomings of our algorithms, and found that the reason for  these apparently superluminal  velocities was  that the field configuration between the two monopoles that are about to merge is such that  it is energetically favourable for a new segment to form in between the two advancing ones.  In some sense, a new segment appears `out of thin air', such that it seems that the monopole has instantaneously moved considerably.  The monopole does not pull from the string and creates string as it moves; instead,   a {\it chunk} of string is formed and the monopole makes a jump forward. 
One way to understand this is to think about the field dynamics in the plane orthogonal to the string that is about to form (see  \cite{Achucarro:1997cx}). Consider the extreme case of a $z$-independent field configuration that is going to become a straight string in the $z$ axis.    Initially the magnetic field in the $x-y$ plane is dispersed over a relatively large area, so its value is below the threshold where our algorithm would recognize the presence of a string, and it is accreting into a smaller area around the $z$ axis --a region with a lower value of the scalar field than the surroundings. At some point the magnetic field grows above the threshold and we see the whole string appear in a single timestep.  

In other words, if the points of windings of $\phi_1$ and $\phi_2$ are almost parallel, it may be favourable for a collection of points to become part of a string all together. Since our simulation only takes snapshots every   20 time steps, this process can remain unnoticed, and the end result is that the segment has apparently grown superluminaly, whereas the reality is  that a new segment has been formed. This can also be understood by considering that the newly formed string segment has its corresponding monopole-antimonopole pair on its ends. The newly formed monopole (say) annihilates with the advancing antimonopole, and the newly formed segment merges with one of the old ones.

It is worth reminding the reader at this stage that we are {\it not} simulating segments and monopoles. We are simulating fields, and the monopoles and segments are consequences of them. Thus, a monopole does not really {\it move}, but it is the movement of the fields which leads to the apparent movement of the monopole. When a new segment forms, the physics of the fields is causal, but the consequence may seem to be a monopole moving ultrarelativistically, when, in essence, it is a different monopole.

Figure~\ref{fig:fast} shows one such situation. The two depicted fragments belong to segments that were behaving normally (i.e., not ultrarelativistically), until they got ready to merge.
When they get close to each other,  the dynamics of the fields leads to the merging as soon as possible: segment ends grow by creating small segments in the ends, and even in between the segments  a new string is formed    (shown as  single points in the figure). Our algorithm is not able to catch these instances, and instead, velocities faster than light are reported.

\begin{figure}[!htbp]
      \includegraphics[width=15cm]{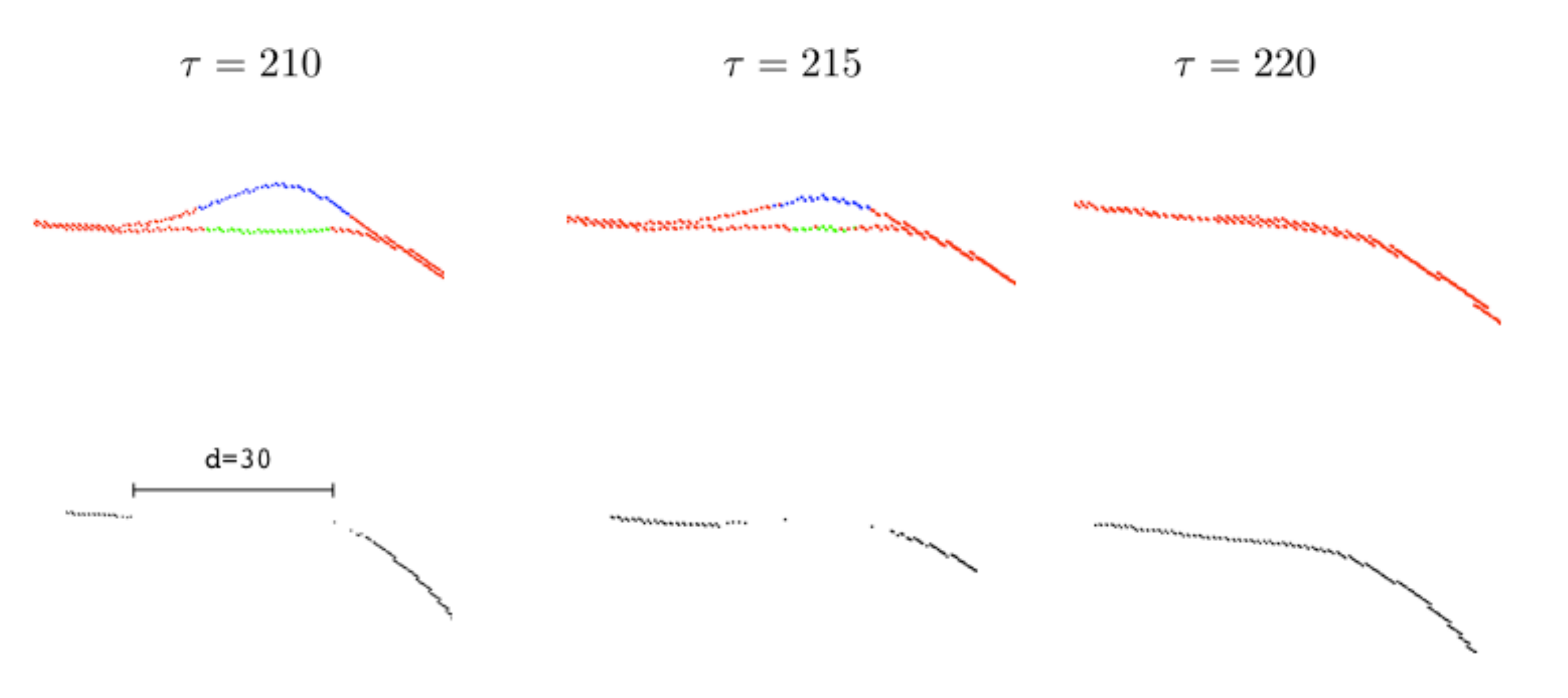} 
     \caption{\label{fig:fast} In this figure we show the merging of two string segments, in what our algorithm would report as   ultrarelativistic velocities. Initially the separation between both string-ends is $d=30$ and in less than $\Delta \tau=10$ both segments have merged. This process can be understood as small string segments being formed both close to the string ends, as well as in between both strings (depicted by those single points in between strings).}
\end{figure}

The question remains now how to automatically factor out those instances, which would clearly corrupt the estimator and give much higher velocities than the true physical one. Learning from our previous experience on segment velocities,  we apply the following  method to  disregard those unphysical velocities: For points that move very fast (in practice we chose velocities  faster than 0.5c),  the velocities obtained by going ``forward" (from  $\tau_1$ to $\tau_2$) and ``backward" (from $\tau_2$ to $\tau_1$) are compared. If   the difference  is higher than $50\%$ of  the value of either of the velocities, we disregard it. We have tested that this is preferable to setting a hard cut-off, because  the results clearly do depend on the cutoffs, and we do not want to artificially enforce that the velocities be sub-luminal in case  this should result from the physics. The monopoles associated to these segment ends are also removed from the analysis.

\section{Numerical results  }\label{results}

In this section we report on the results obtained from our simulations, after carefully accounting for the caveats described in the previous section.

We first checked that the simulations reached the scaling regime.  Scaling can be checked  by the evolution of the energy of the system, as mentioned earlier, by checking that $T_{00}^{-1/2}$ is linear with time $\tau$. This is a measure that depends  only on the values of the fields as simulated, without further treatment. We show the scaling on our simulations using the energy as a measure of scaling in Fig.~\ref{t00}.

\begin{figure}[!htbp]
\centering
      \includegraphics[width=8.5cm]{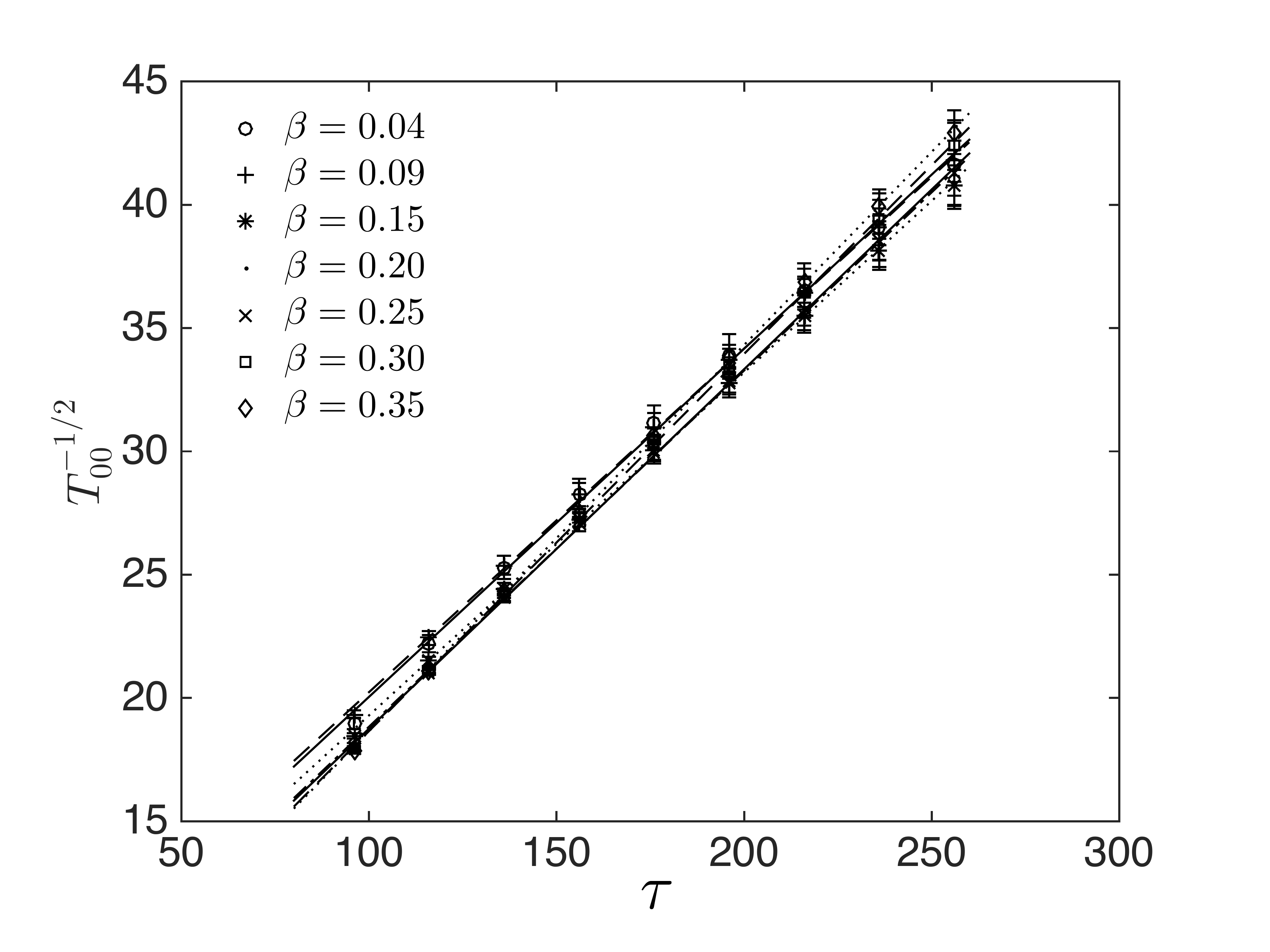} 
      \includegraphics[width=8.5cm]{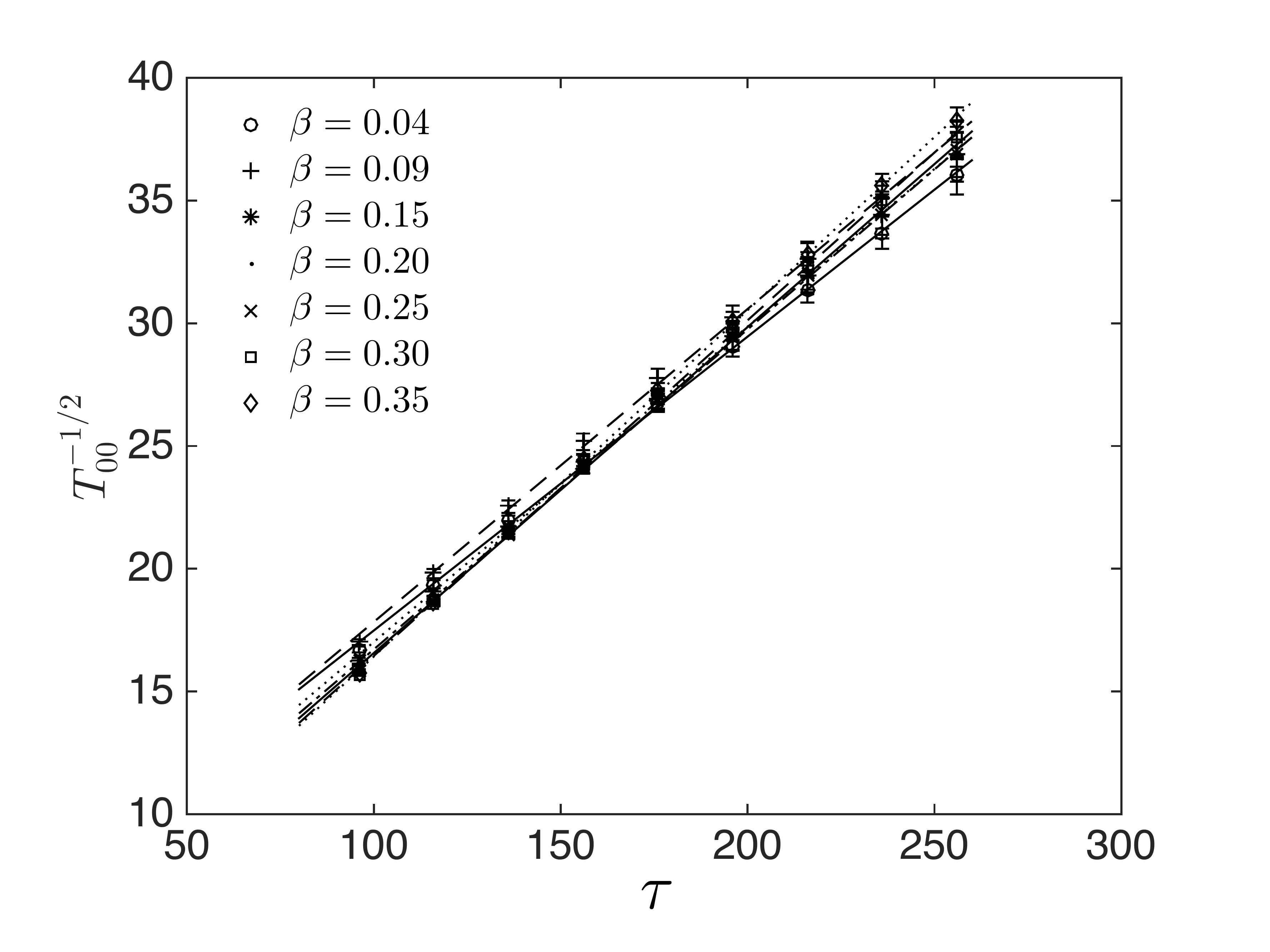}
     \caption{\label{t00} This figure shows that our simulations reach a scaling regime for all $\beta$. The left panel shows the case of radiation domination, the right panel the case of matter domination.}
\end{figure}

There are also two other diagnostics for scaling that depend on the total length of string  $\mathcal{L}$ and the total number of monopoles $\mathcal{N}$. One can obtain a  VOS-type length scale by the following combinations $\sqrt{\frac{V}{\mathcal{L}}}$ and on $\left(\frac{V}{\mathcal{N}}\right)^{1/3}$. These quantities are derived, meaning that one has to extract this information from the simulation using estimators, and therefore, in some sense, they are more indirect than  using the energy to check for scaling. We show the behaviour of these quantities in Figure~\ref{fig:scaling} for two extreme cases (radiation and $\beta=0.04$) and (matter and $\beta=0.35$).  These curves show that the system does indeed reach scaling fairly early. The slope of the curves related to   $\mathcal{L}$ and  $\mathcal{N}$  is what was used above  Eq. (\ref{params}) to determine the VOS-type scaling parameters. The values of $\gamma_{\mathcal{L}}$ and $\gamma_{\mathcal{N}}$ can be found in Table~\ref{tab-gamma} for every $\beta$ studied.

\begin{figure}[htbp]
   \centering
   \includegraphics[width=8.5cm]{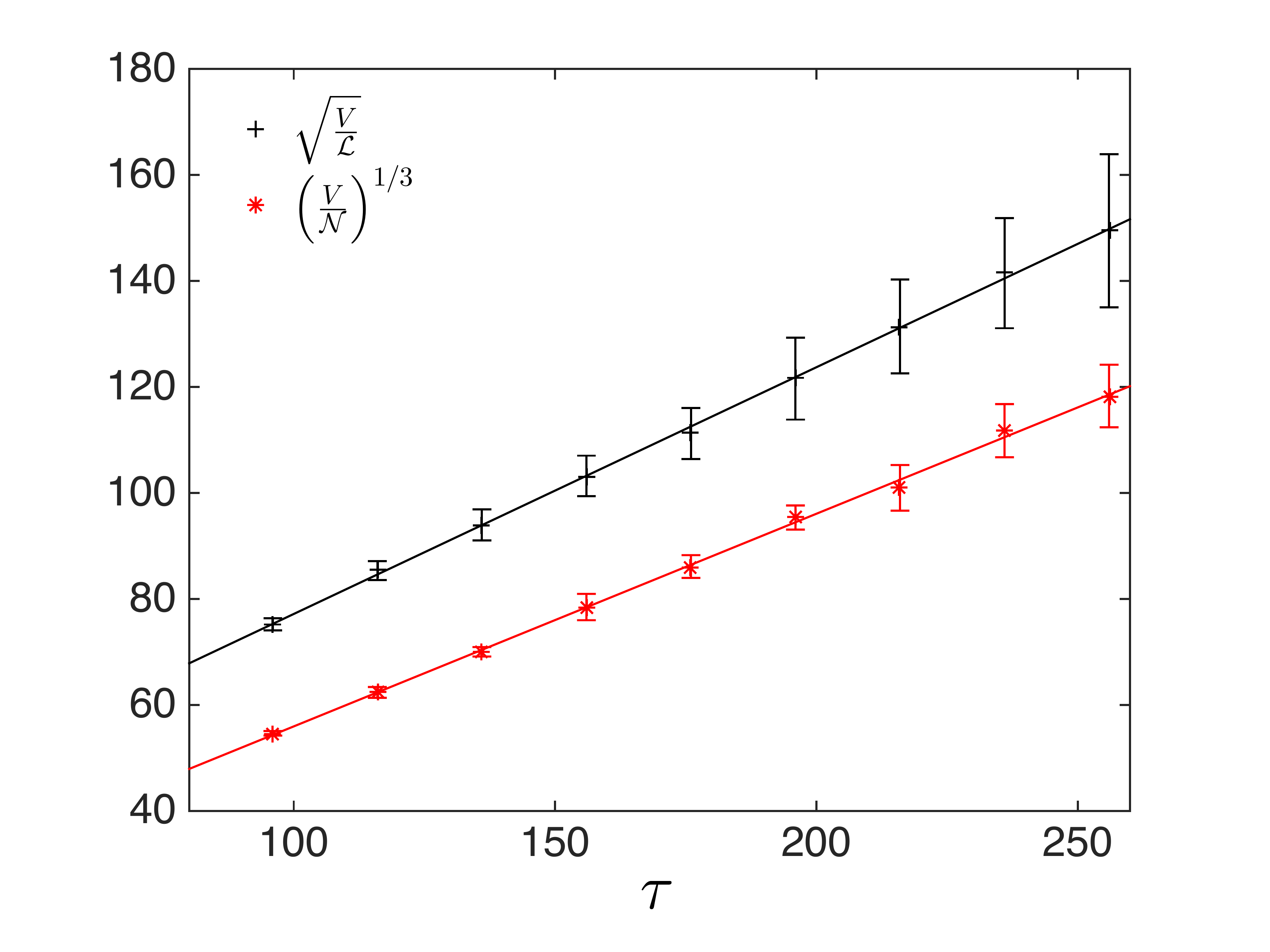} 
   \includegraphics[width=8.5cm]{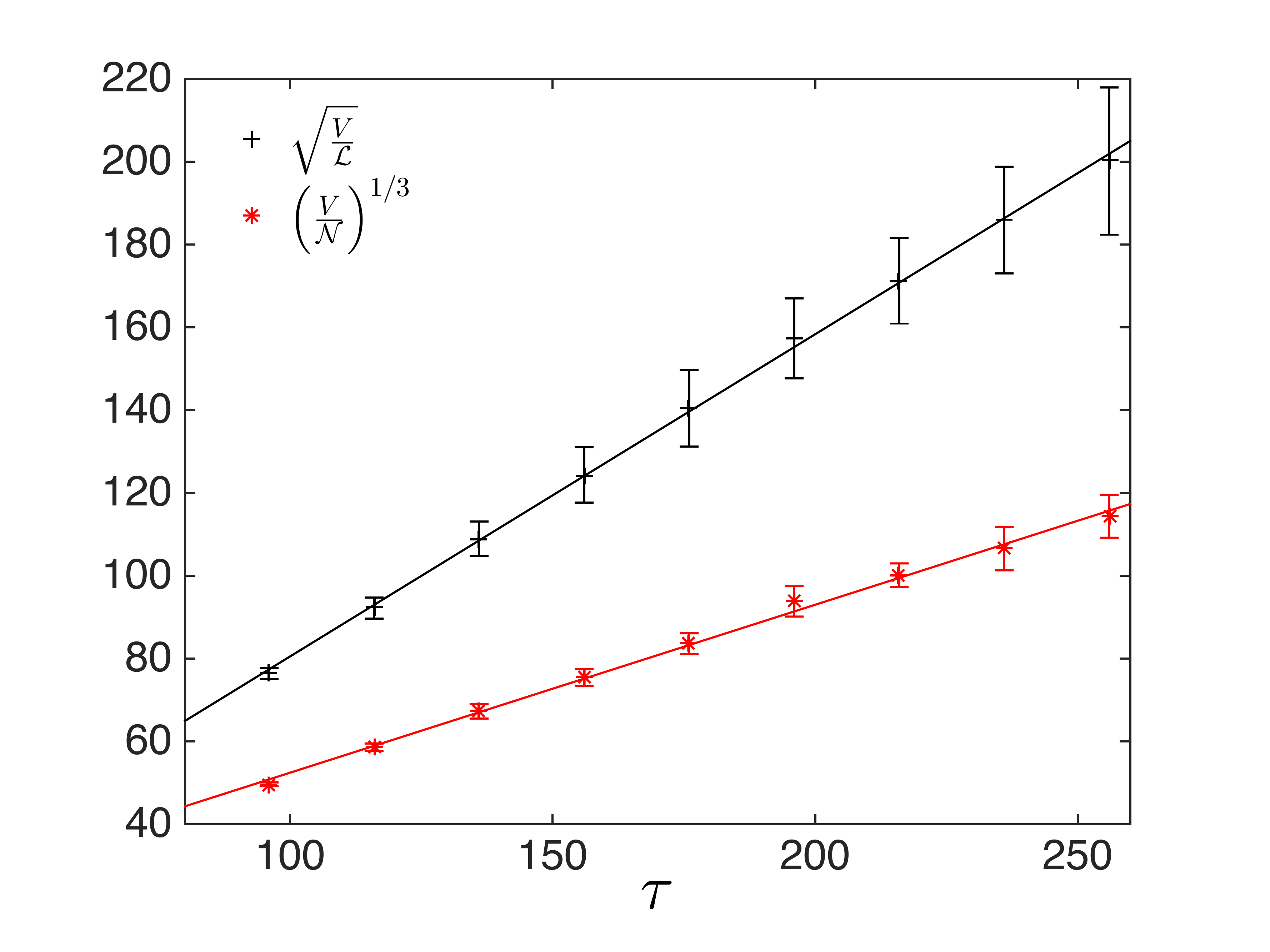}
   \caption{Scalings in total string length and monopole number for $\beta=0.15$ in radiation era (left) and for $\beta=0.35$ in matter era (right).}
   \label{fig:scaling}
\end{figure}

\begin{table*}
\begin{center}
\begin{tabular}{|c|c|c||c|c|}
\hline
 & \multicolumn{2}{c||}{Radiation} & \multicolumn{2}{c|}{Matter} \\
 \hline
 $\beta$ & $\gamma_{\mathcal{L}}$ & $\gamma_{\mathcal{M}}$ & $\gamma_{\mathcal{L}}$ & $\gamma_{\mathcal{M}}$ \\  \hline

 0.04 & 0.38 $\pm$ 0.05 & 0.48 $\pm$ 0.04 & 0.28 $\pm$ 0.02 & 0.41 $\pm$ 0.02  \\  \hline
 
 0.09 & 0.40 $\pm$ 0.04 & 0.43 $\pm$ 0.02 & 0.33 $\pm$ 0.02 & 0.38 $\pm$ 0.02  \\  \hline
 
 0.15 & 0.47 $\pm$ 0.09 & 0.40 $\pm$ 0.02 & 0.40 $\pm$ 0.04 & 0.35 $\pm$ 0.03  \\  \hline
 
 0.20 & 0.52 $\pm$ 0.08 & 0.41 $\pm$ 0.04 & 0.46 $\pm$ 0.04 & 0.37 $\pm$ 0.02  \\  \hline
 
 0.25 & 0.61 $\pm$ 0.09 & 0.42 $\pm$ 0.05 & 0.54 $\pm$ 0.06 & 0.37 $\pm$ 0.02  \\  \hline
 
 0.30 & 0.81 $\pm$ 0.09 & 0.46 $\pm$ 0.05 & 0.65 $\pm$ 0.09 & 0.37 $\pm$ 0.03  \\  \hline
 
 0.35 & 1.06 $\pm$ 0.09 & 0.49 $\pm$ 0.05 & 0.78 $\pm$ 0.09 & 0.41 $\pm$ 0.03  \\  \hline
 
\end{tabular}
\caption{\label{tab-gamma} Values of the VOS-type length estimators for the total length of segments $\gamma_\mathcal{L}$ and the total number of monopoles $\gamma_{\mathcal{M}}$ (\ref{params}) in the box for every $\beta$. }
\end{center}

\end{table*}

One of the main results in this work is the  values of the velocities of the semilocal network, both for strings and for monopoles, using our different estimators. In  Table~\ref{veldirect} we show the velocity values for every $\beta$ in the radiation and matter dominated epochs obtained by following the positions of each string segment and monopole, and averaging over all cases (the errors given are statistical errors). Overall one can see that the velocities in radiation are somewhat higher than the velocities in matter, as expected due to a lower damping term in the former epoch. More interestingly, there does not seem to be a strong dependency of the velocities on $\beta$; indeed, note that all these values are equivalent within their statistical errors.

In Table~\ref{velfield} we show the velocities obtained by local field estimators, cf. Eq. (\ref{Markestimators}). Note that these estimators do not distinguish between strings and monopoles, and give one number for the average network velocity for semilocal strings as a whole. Once again, we obtain that velocities during radiation are faster than in matter; and we also note that both local field estimators give equivalent velocities. In this case  there seems to be a trend, and velocities seem to decrease for increasing $\beta$.

\begin{table*}
\begin{center}
\begin{tabular}{|c|c|c||c|c|}
\hline
 & \multicolumn{2}{c||}{Radiation} & \multicolumn{2}{c|}{Matter} \\
 \hline
 $\beta$ & $v_{\mathcal{L}}$ & $v_{\mathcal{M}}$ & $v_{\mathcal{L}}$ & $v_{\mathcal{M}}$ \\  \hline

 0.04 & 0.345 $\pm$ 0.010 & 0.574 $\pm$ 0.010 & 0.266 $\pm$ 0.010 & 0.505 $\pm$ 0.010  \\  \hline
 
 0.09 & 0.338 $\pm$ 0.010 & 0.583 $\pm$ 0.010 & 0.265 $\pm$ 0.010 & 0.510 $\pm$ 0.010  \\  \hline
 
 0.15 & 0.337 $\pm$ 0.010 & 0.600 $\pm$ 0.012 & 0.262 $\pm$ 0.011 & 0.509 $\pm$ 0.010  \\  \hline
 
 0.20 & 0.337 $\pm$ 0.010 & 0.591 $\pm$ 0.010 & 0.260 $\pm$ 0.010 & 0.519 $\pm$ 0.010  \\  \hline
 
 0.25 & 0.337 $\pm$ 0.010 & 0.591 $\pm$ 0.010 & 0.261 $\pm$ 0.010 & 0.520 $\pm$ 0.010  \\  \hline
 
 0.30 & 0.342 $\pm$ 0.014 & 0.596 $\pm$ 0.010 & 0.259 $\pm$ 0.010 & 0.524 $\pm$ 0.010  \\  \hline
 
 0.35 & 0.337 $\pm$ 0.013 & 0.600 $\pm$ 0.012 & 0.262 $\pm$ 0.011 & 0.521 $\pm$ 0.012  \\  \hline
 
\end{tabular}
\caption{\label{veldirect} Average velocities   for   segments  ($v_{\mathcal{L}}$) and monopoles ($v_{\mathcal{M}}$) obtained by following the positions of the segments  and the monopoles during the simulation. The errors are statistical errors obtained by averaging. The velocities in radiation are noticeably higher than in matter, but there does not seem to be a strong dependence on $\beta$ for the velocities.}
\end{center}
\end{table*}

\begin{table*}
\begin{center}
\begin{tabular}{|c|c|c||c|c|}
\hline
 & \multicolumn{2}{c||}{Radiation} & \multicolumn{2}{c|}{Matter} \\
 \hline
 $\beta$ & $v_{F}$ & $v_{G}$ & $v_{F}$ & $v_{G}$ \\  \hline

 0.04 & 0.57 $\pm$ 0.02 & 0.55 $\pm$ 0.02 & 0.41 $\pm$ 0.01 & 0.41 $\pm$ 0.01  \\  \hline

 0.09 & 0.57 $\pm$ 0.02 & 0.56 $\pm$ 0.02 & 0.42 $\pm$ 0.01 & 0.41 $\pm$ 0.01  \\  \hline
 
 0.15 & 0.55 $\pm$ 0.02 & 0.55 $\pm$ 0.01 & 0.41 $\pm$ 0.01 & 0.41 $\pm$ 0.01  \\  \hline
 
 0.20 & 0.53 $\pm$ 0.02 & 0.53 $\pm$ 0.02 & 0.40 $\pm$ 0.01 & 0.40 $\pm$ 0.01  \\  \hline
 
 0.25 & 0.51 $\pm$ 0.02 & 0.51 $\pm$ 0.02 & 0.39 $\pm$ 0.01 & 0.39 $\pm$ 0.01  \\  \hline
 
 0.30 & 0.49 $\pm$ 0.02 & 0.50 $\pm$ 0.02 & 0.38 $\pm$ 0.01 & 0.38 $\pm$ 0.01  \\  \hline
 
 0.35 & 0.47 $\pm$ 0.02 & 0.48 $\pm$ 0.02 & 0.36 $\pm$ 0.02 & 0.37 $\pm$ 0.02  \\  \hline
\end{tabular}
\caption{\label{velfield}  Velocities obtained from local field estimators, Eq. (\ref{Markestimators}), for all $\beta$ and radiation and matter dominations. Velocities in radiation domination are higher than in matter domination. Now there is moderately robust statistical evidence that the velocity of the network decreases for increasing $\beta$. }
\end{center}
\end{table*}

The procedure of following the positions of the string segments and of the monopoles does not only yield a network velocity, but it also allows us to obtain the velocity of each segment and monopole. We can thus follow the history of each segment and monitor whether they merge with other segments, disappear, or just continue to evolve in the network. This is very interesting information from the point of view of the VOS model, because we can now tell if there is a correlation between string segment size and its velocity, for example.

\begin{figure}[!htbp]
   \centering
   \includegraphics[width=5.9cm]{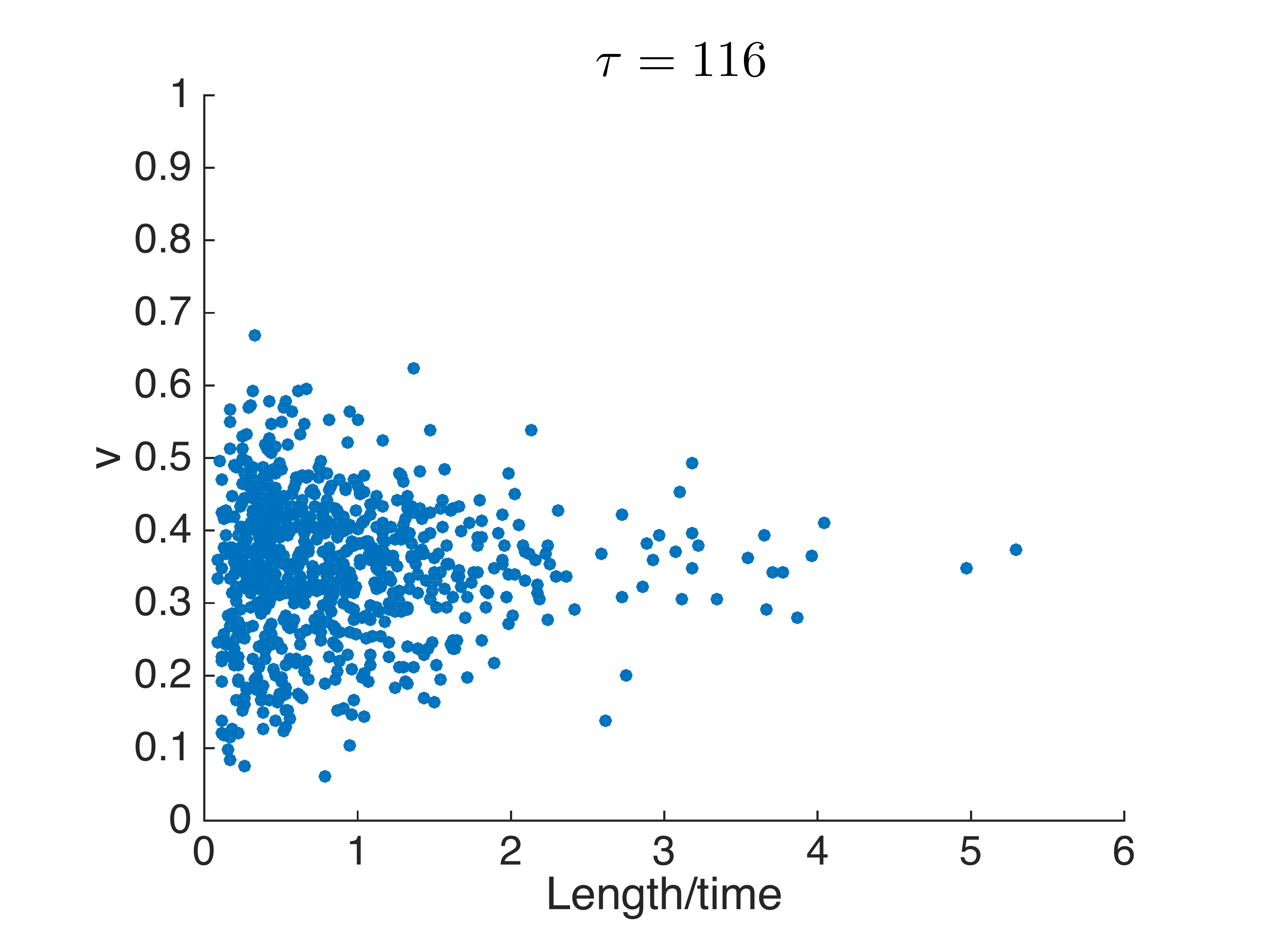} 
   \includegraphics[width=5.9cm]{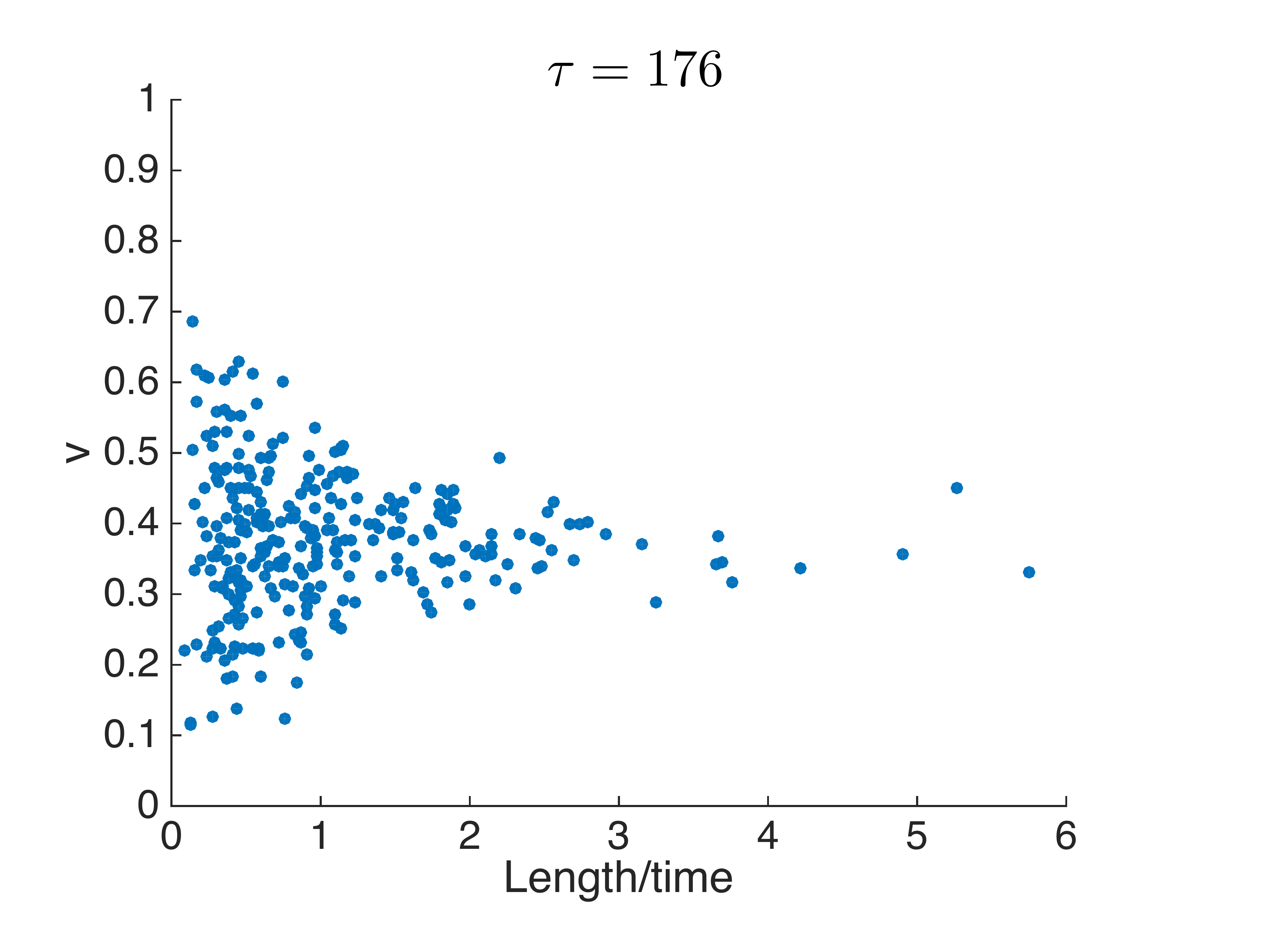} 
   \includegraphics[width=5.9cm]{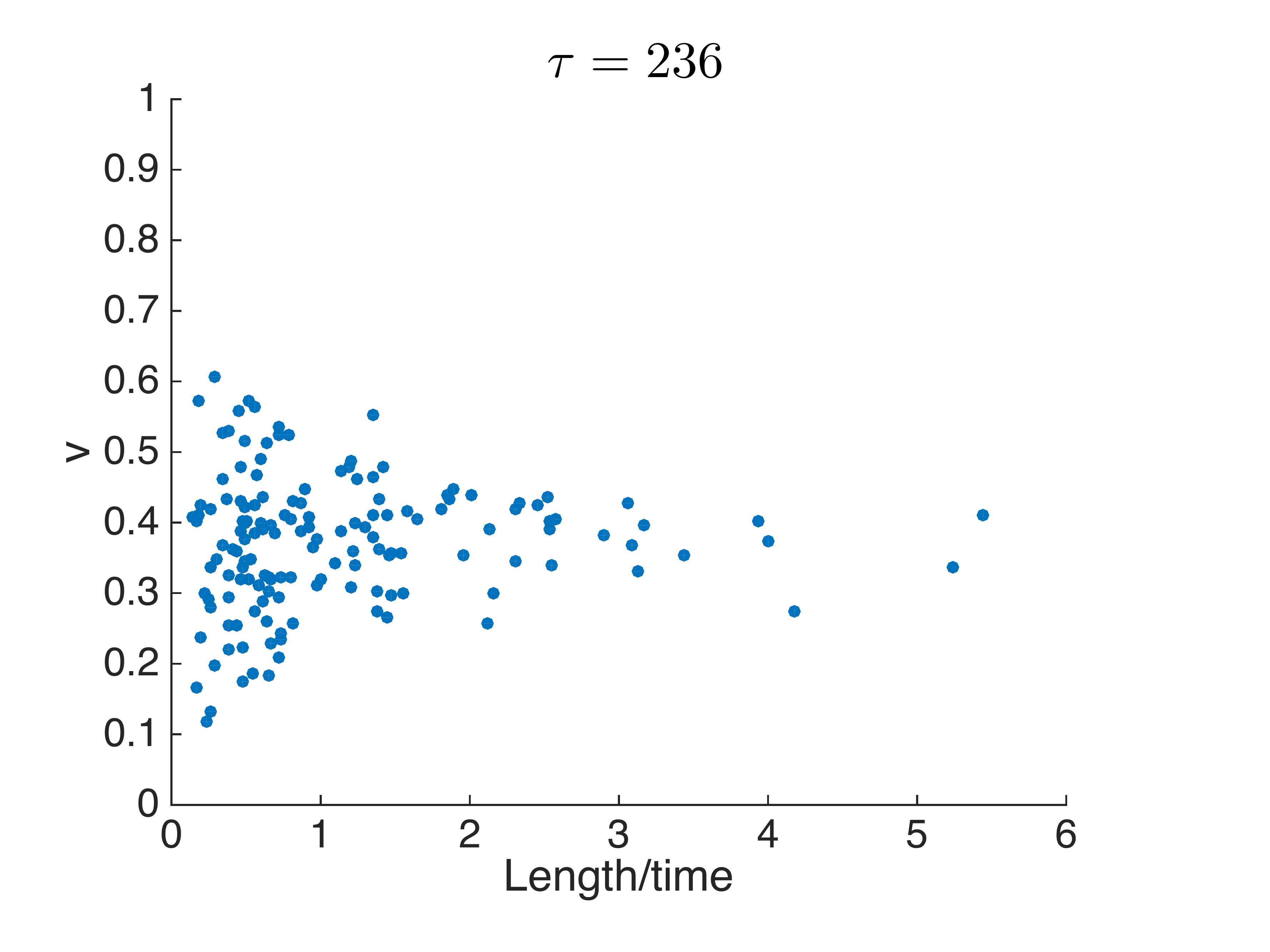} 
   \caption{This plot shows the distribution of segments with respect to velocity for simulation in radiation domination and $\beta=0.04$  for all seven simulations.  Each point represents a segment in the network, where in the x-axis the length of the segment divided by time is shown and its velocity in the y-axis. This is the case where segments {\it flow} through the network, i.e., they do interact with any other segment in the next time step.
   \label{fig:point-velocityf}}
\end{figure}

\begin{figure}[!htbp]
   \centering
   \includegraphics[width=5.9cm]{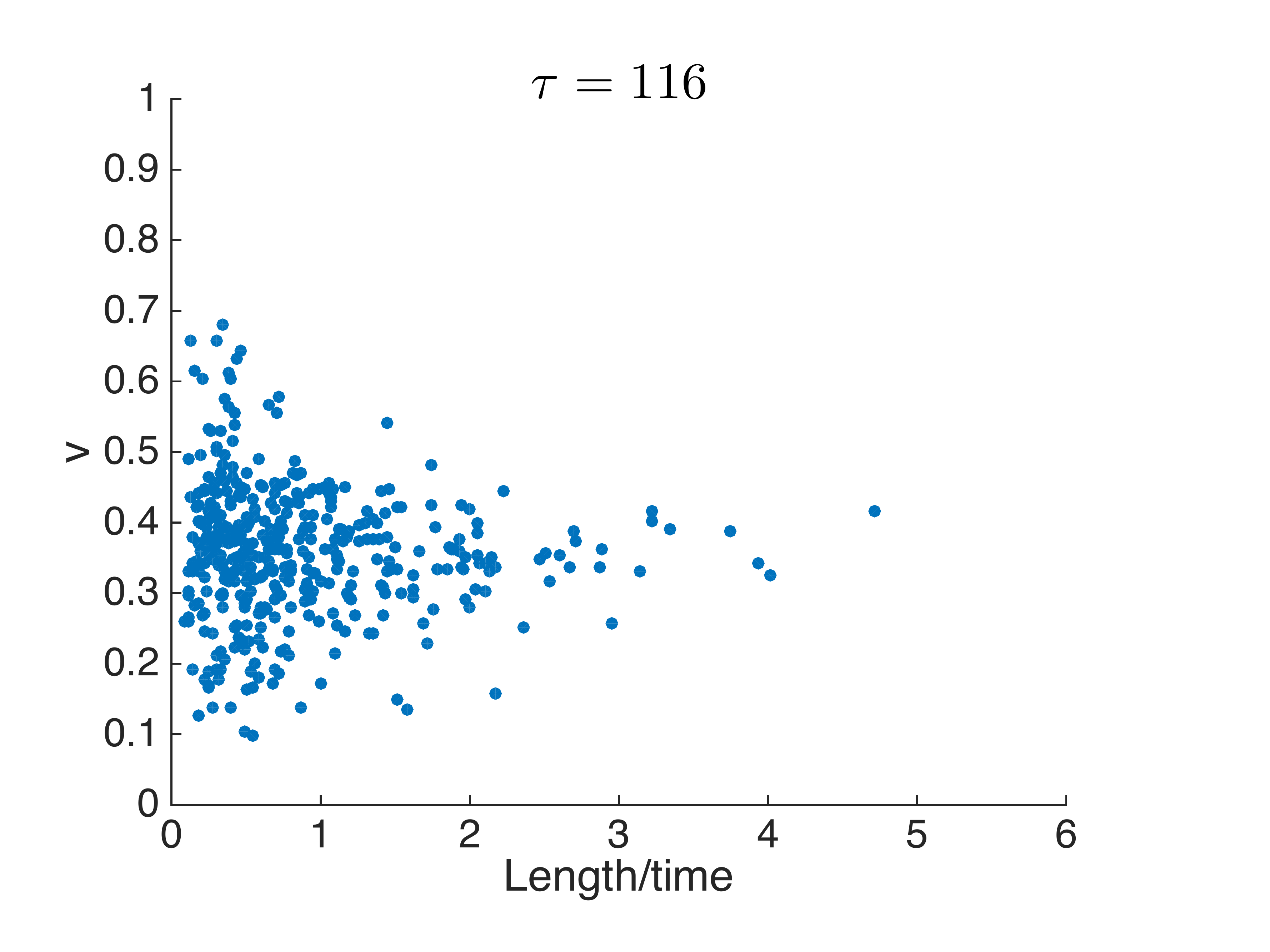} 
   \includegraphics[width=5.9cm]{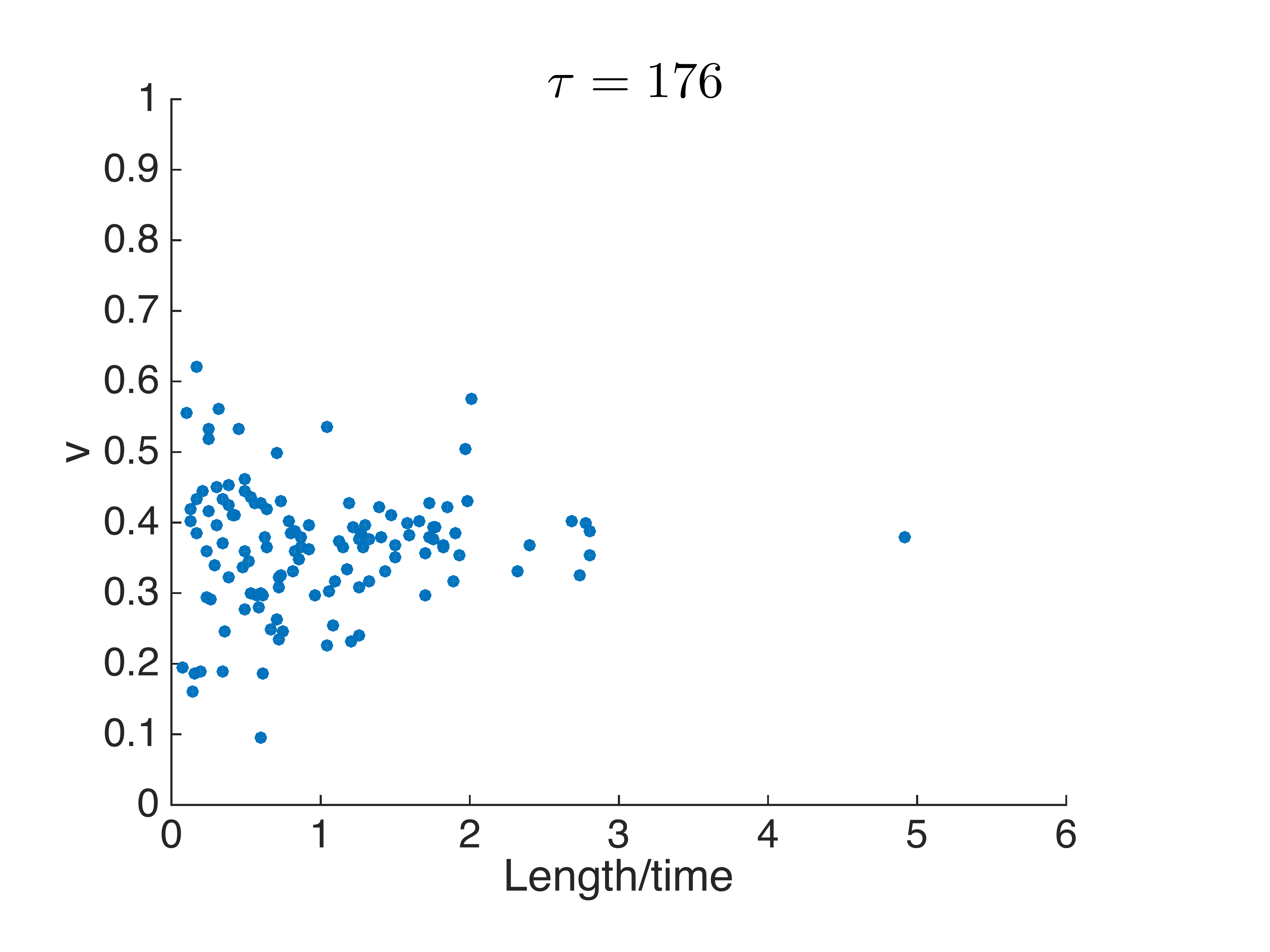} 
   \includegraphics[width=5.9cm]{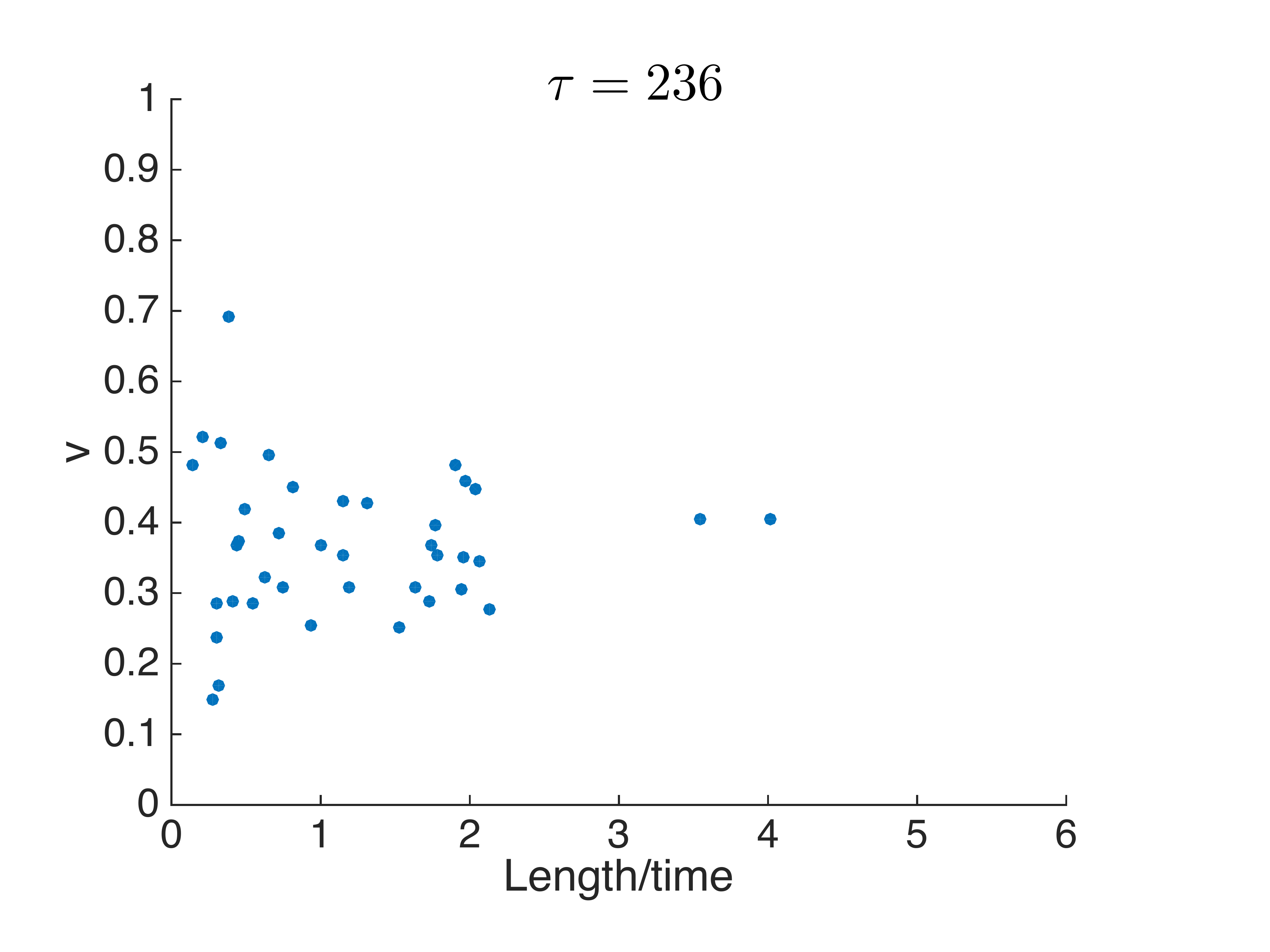} 
   \caption{This figure is similar to  Fig.~\ref{fig:point-velocityf}, but in this case the segments that are plotted are those that {\it merge} with other segments before the next time step.
   \label{fig:point-velocitym}}
\end{figure}

In the subsequent figures we only report the results for $\beta=0.04$ and radiation domination. The situation with other $\beta$ and matter domination is analogous; figures for other values of $\beta$ are provided as Supplemental Material \cite{supplemental} to this paper. The choice of $\beta=0.04$ to show the results in the main body of the paper was made since the number of segments for this $\beta$ is higher, and thus the different phenomena can be more easily appreciated.

We show in Figs.~\ref{fig:point-velocityf} and \ref{fig:point-velocitym}  the velocity versus length distribution for some time-steps. In these figures, each point corresponds to a segment, and these are distributed according to their length divided by time (x axis) and velocity (y axis) for three different times ($\tau=$116, 176, 236).  These segments have been divided into two groups according to their behaviour {\bf in the next time-step}: On the one hand, in Fig.~\ref{fig:point-velocityf} we plot the segments  which do not interact with other string segments in the next time step, i.e., segments which {\it flow} through the network. On the other hand, in Fig.~\ref{fig:point-velocitym} we show the segments that do merge with other segments before the next time step.

It is worth mentioning that the points are scattered around a central value, which is roughly the same for all times and all lengths, specifically $v\sim0.35$. The scatter is larger around short segments, mainly because there are many more short than long segments. This is rather clear in the {\it flow} case, because there are more segments in this case than in the {\it merge} case; though we could say that this is generic for all cases.

This information can be shown also, maybe in a more compact way,  in the form of histograms; these are in some sense the marginalized distribution functions of   Figures~\ref{fig:point-velocityf} and \ref{fig:point-velocitym}. For example, in the top panel of Fig.~\ref{histvelseg} we show the distribution of the lengths divided by time of the number of segments during the evolution for $\beta=0.04$ in radiation domination. The segments are binned in 10 bins with uniform width, and we show together three cases: in blue we depict the segments which are {\it flowing}, in green the ones that are {\it merging} and in yellow we show the segments that {\it collapse}. Remember that those behaviours are determined only for the reported time-step, i.e., a segment that {\it flows} at a given time-step can be {\it merging} on the next one. Note that in the last time step we have no information whether the segments will flow, merge or collapse, so we just choose to show them as flow segments. In the bottom panel we show the same histogram, but instead of showing the number of segments in each bin, we also account for the length-per-time of the strings; we add up the total length-per-time of the segments in each bin, and show that length-per-time in each bin.

The histograms with information about velocities can be seen in Fig.~\ref{monopole-velocity}. The top panel shows the distribution of number of segments per velocity (for $\beta=0.04$ in radiation domination, as before). The bottom panel shows the corresponding distribution for monopoles. These velocities are computed instantaneously, i.e., they are the velocities obtained by measuring the distance traveled by monopoles between two adjacent time-steps (not by averaging over the whole history of the monopole). Note that there is no information for time $\tau=256$ because we have no  ``next time step" to compute the velocity with. The color code is the same is in the previous figure: in blue we depict the strings (segments or monopoles, correspondingly) which are   flowing, in green the ones that are  merging and in yellow we show the strings that {\it collapse} before the next time step.

\begin{figure}[!htbp]
   \centering
   \includegraphics[width=15cm]{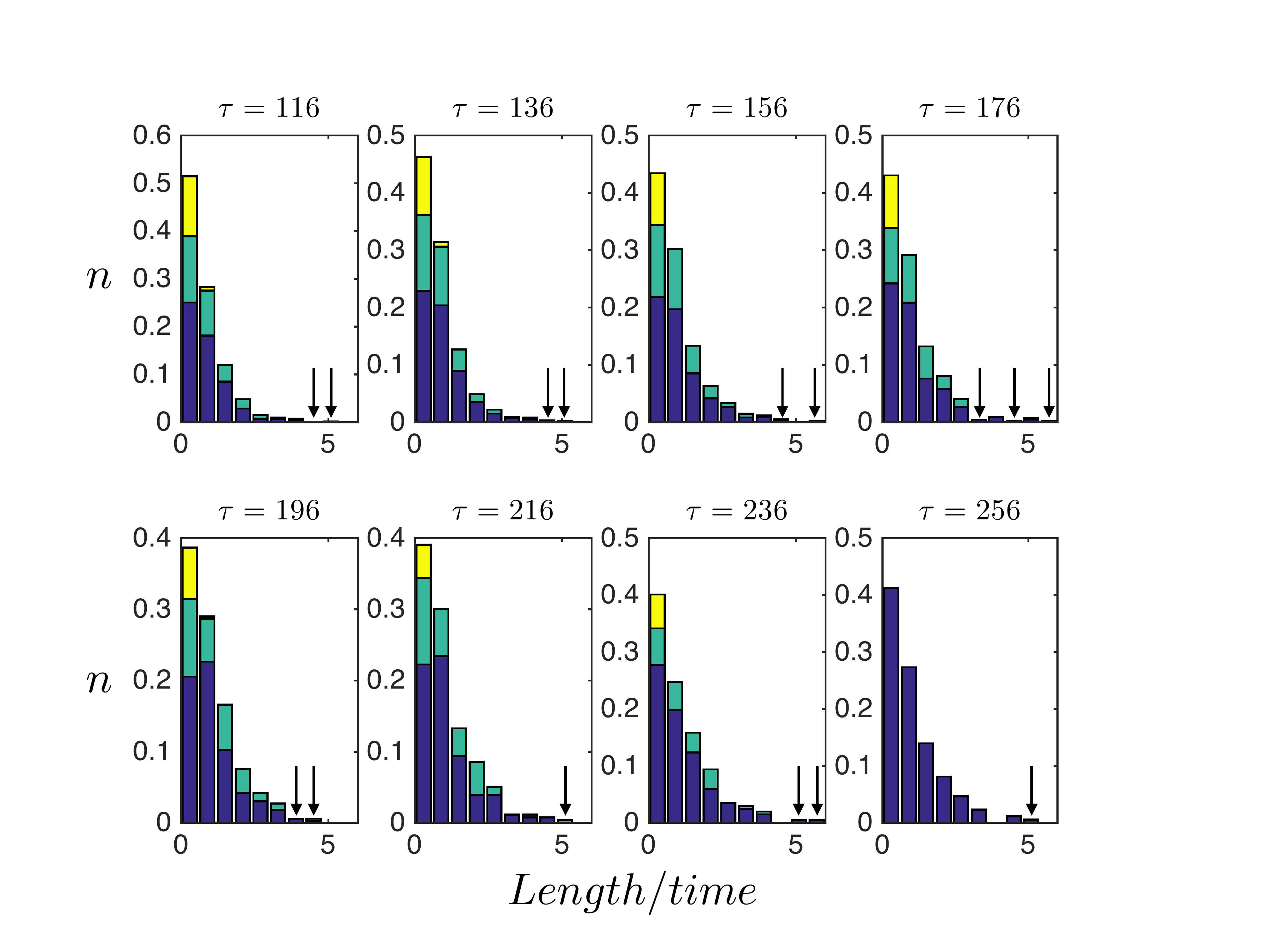}
 \includegraphics[width=15cm]{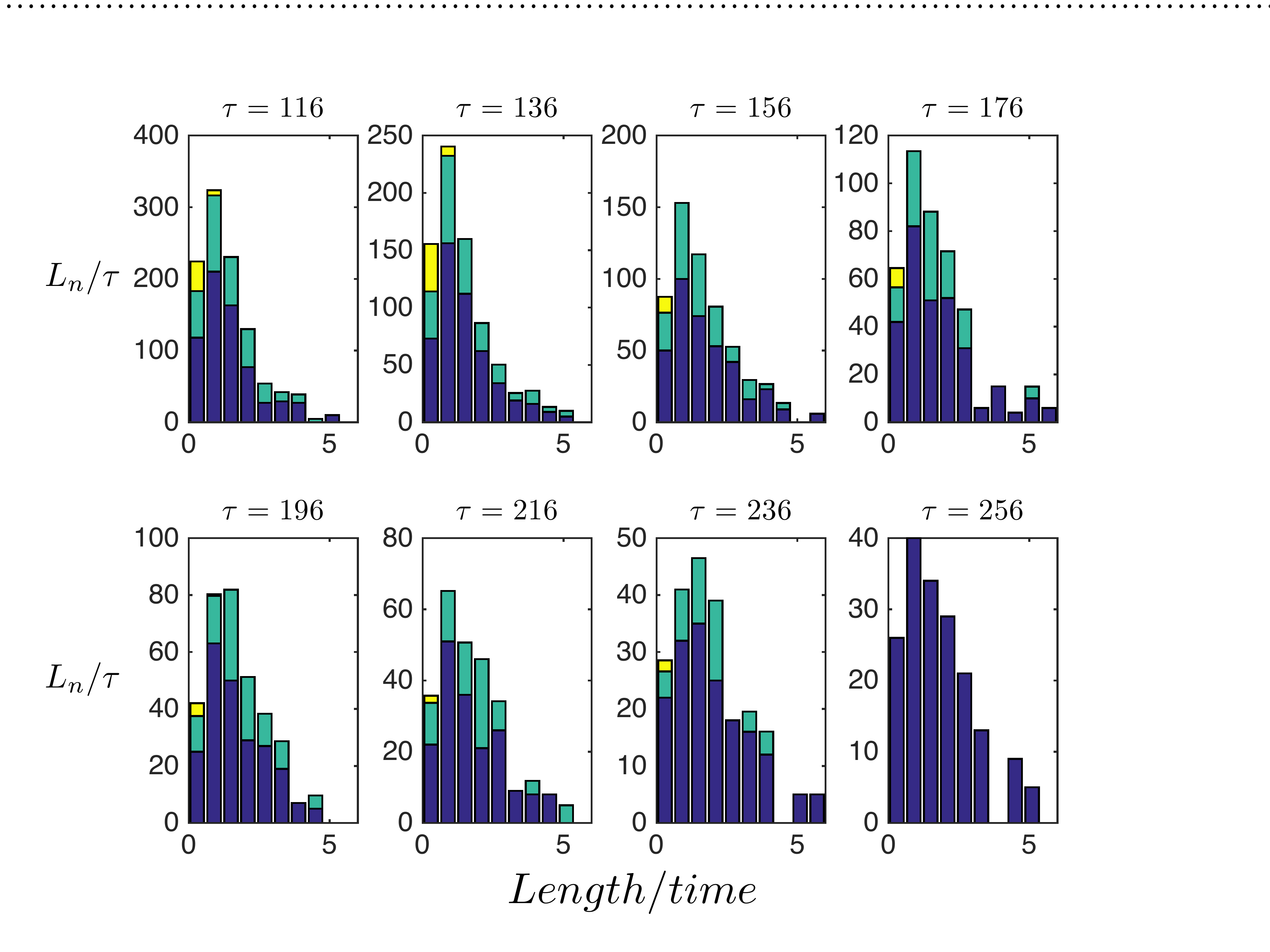}
   \caption{These histograms show the distribution of the segments during the network evolution for $\beta=0.04$ in the radiation dominated era for all seven simulations, where the segments are binned in 10 bins with uniform width. The top panels show the fractional distribution of the number of segments $n$ with their lengths divided by time, whereas in the bottom panels show the analogous distribution for the total lengths divided by time $L_n/\tau$.  The colors represent different types of segments, depending on their future behaviour: in blue  segments which are {\it flowing}, in green segments that are {\it merging} and in yellow   segments that are {\it collapsing} before the next time-step. We write arrows to remark that in those instances, there are a few (one or two) segments in that bin,  which are hard to see in the top panels, but can be seen in the bottom ones.
 Note that in the last time step we have no information whether the segments will flow, merge or collapse, so we just choose to show them as flow segments.
   \label{histvelseg}}
\end{figure}

The histograms are consistent with a scaling behaviour of the segment distributions. The length distributions show that, logically, only short segments collapse before the next time step. However, mergings happen for all string lengths. There  is no preference on the velocity of the collapsing segments: there are collapsing segments with a wide variety of velocities. The velocities follow approximate Gaussian distributions, centered in the average velocity, corresponding  to the velocities of Table~\ref{veldirect}.   
 
The monopole velocity histograms also show a (wider) Gaussian-like distribution, with instances in which monopoles approach $v=1$ (though this may be an artefact of our algorithms, as explained in section~\ref{superluminal}). It is easy to convince oneself that the averages of these velocities agree with the ones in Table~\ref{veldirect}.  It is worth noticing that the velocities of monopoles that are about to annihilate are rather high: most of the times they are higher than the average velocities of the segments. Therefore, even though the segment velocities for collapsing segments show no evidence of a bias, the monopole velocities do. This could be understood by realizing that the monopole velocity only takes into account the string-ends, whereas in the segment velocity, the whole segment contributes, and apparently only the segment ends get high velocities in collapse.

\begin{figure}[!htbp]
   \centering
   \includegraphics[width=15cm]{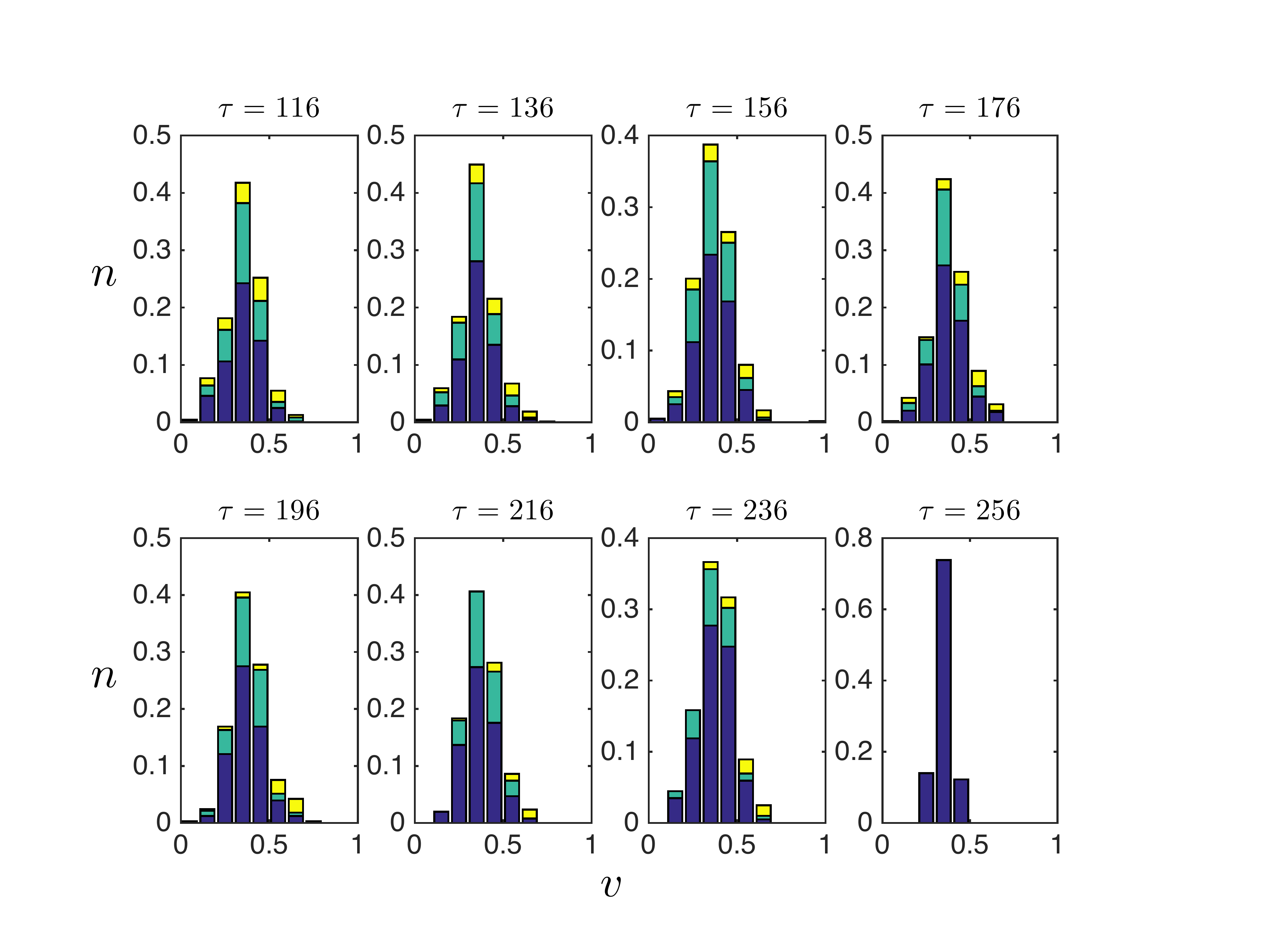} 
    \includegraphics[width=15cm]{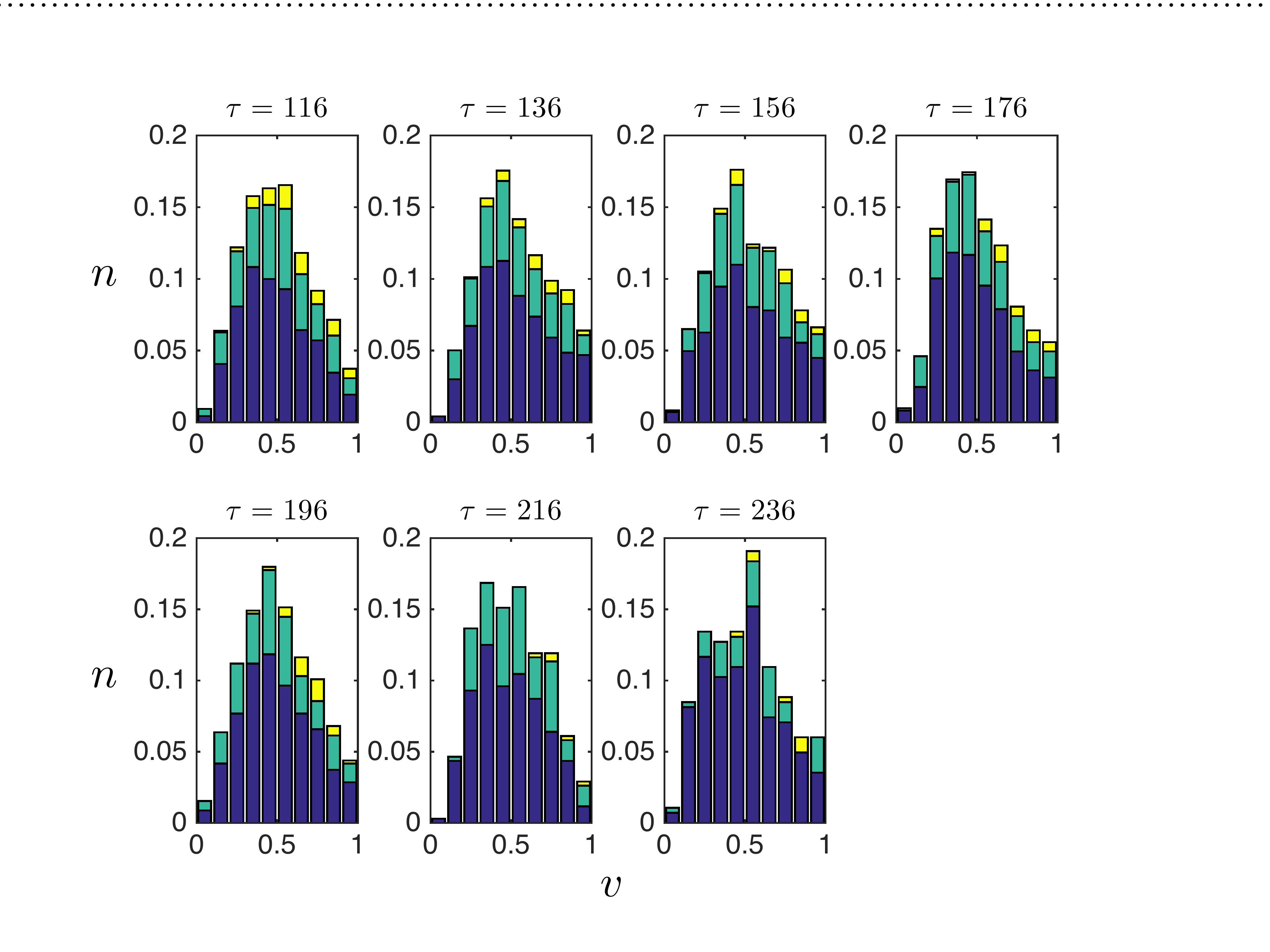} 
   \caption{These histograms show the velocity $v$ distribution of the segments (top) and monopoles (bottom) during their evolution, for radiation and $\beta=0.04$. The velocities are binned in 10 bins with uniform width.   The color code is analogous to that of Fig. \ref{histvelseg}: blue corresponds to strings that do not interact with other strings before the next time-step (flow),  green is for strings which merge with other segments and yellow is for strings that disappear before the next time-step because the segment collapses. Note that in the last time step of segment velocities we have no information whether the segments will flow, merge or collapse, so we just choose to show them as flow segments. Note also that  unlike in the segment case, where the velocity has been calculated {\it forward} and {\it backward}, for monopoles we have only calculated the velocity {\it forward}, and therefore we  do not have information to compute the velocity at the last time step. \label{monopole-velocity}}
\end{figure}

Finally, it is interesting to investigate what  the pattern of merging of different segments is. In order to do so, we chose one of the largest segments at the last time-step for $\beta=0.04$, and traced back its history to see what were its ``constituents". This information can be found in Fig.~\ref{b04family}, in the form of a ``family tree". This shows that at   very early times, many mergings happen. This does not mean that many segments have joined at the same time; most likely the segments have been joining by pairs, but our choice of time steps to measure the network is too coarse to distinguish all these mergings. The sum of the length of the constituents does not match the final segment; clearly, this is because segments can grow or shrink in their evolution. Also, there are some segments that remain quite solitary for most of their life. This exercise of following the family history of a segment highlights once more the complicated dynamics of  semilocal string networks.

\begin{figure}[htbp]
   \centering
   \includegraphics[width=17cm]{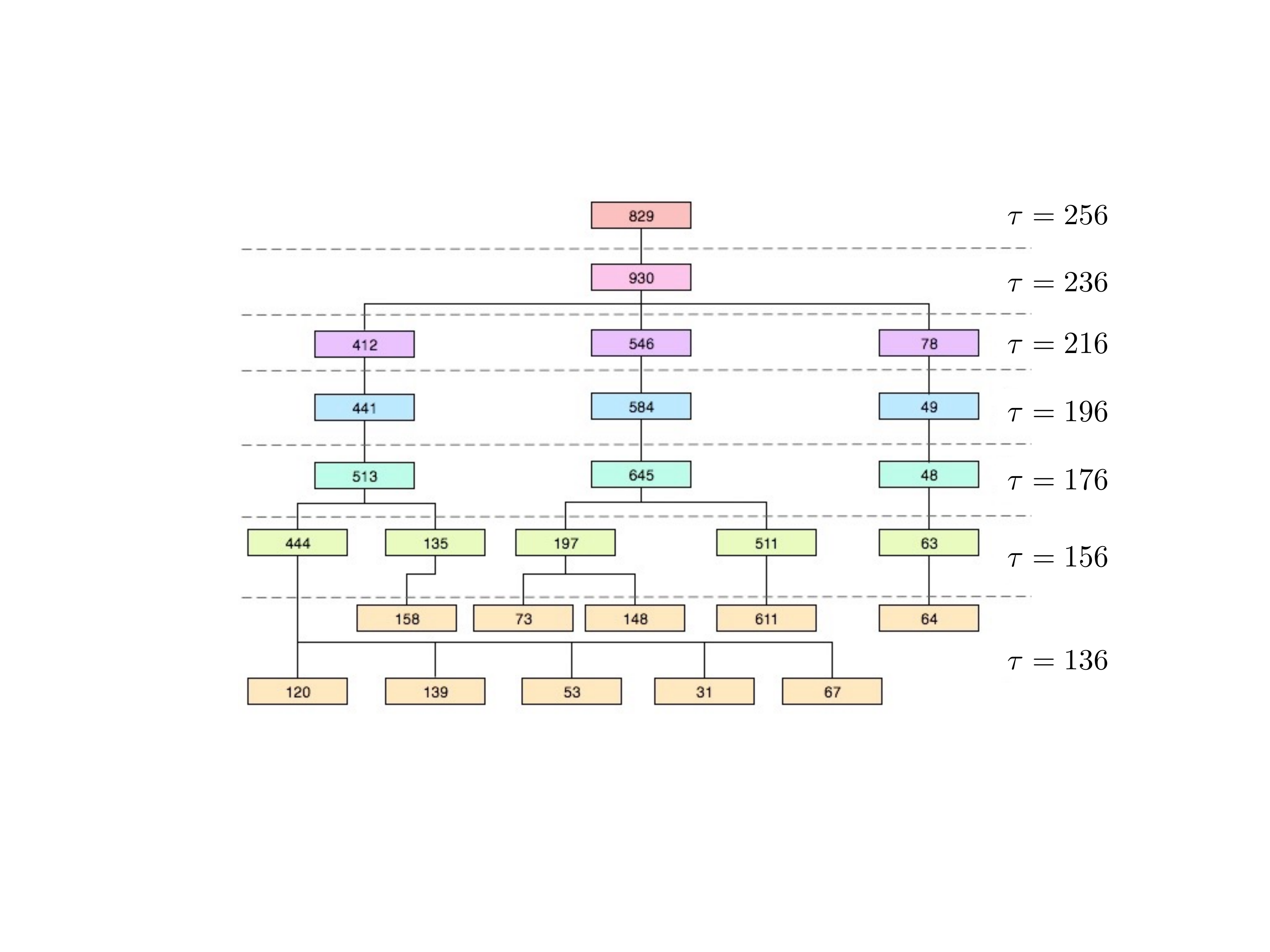} 
   \caption{The ``family tree" of a segment for $\beta=0.04$ in radiation. The number inside the box denotes the length of each segment, and time runs upwards.}
   \label{b04family}
\end{figure}

\section{Conclusions and discussion}

In this work we have further investigated the evolution of networks of semilocal strings using field theoretical simulations. We have estimated the length and velocity of the semilocal strings (including the string ends, which can be understood as global monopoles) using different estimators, coming from both direct field theory based diagnostics and from identifications of the position of the strings.  As well as characterizing the network of defects, these measures are indispensable to obtain a VOS-type effective model for semilocal strings. Before this work, the velocities of   semilocal strings (both as a network and as individual segments, as well as the monopoles) were unknown.

The VOS-type length estimators for string segments and monopoles (\ref{params}) can be found in Table~\ref{tab-gamma}. These values had already been computed in a previous work \cite{Achucarro:2013mga} using a different technique. In \cite{Achucarro:2013mga} the semilocal string segments were defined as collections of  adjacent points with a magnetic field higher than some threshold, and that volume of points was divided by an estimate of the cross-section of the strings to get a length. In this work we repeated this analysis and compared it to the new length estimator, which is based on following the points with windings in the simulation. We found that these two length estimators do not match, and that there is roughly a factor ~1.5 difference between the two length estimators. We identified the source of error as a numerical error on the estimation of the string cross-section, and therefore the values for the length estimators that we obtain in the present work can be considered as corrections to the ones in  \cite{Achucarro:2013mga}. However, the differences obtained lie within  the 1-$\sigma$ uncertainties. It should also be noted that the length corresponding to the monopoles $\gamma_{\mathcal{N}}$ is roughly equivalent in both works, since the number of segments (monopoles) does not depend on the error corresponding to the cross-section of the strings.

One big advantage of being able to characterize the semilocal strings by the points (plaquettes) with winding is that we get a one-dimensional description for the position of the strings. This allows us to follow string segments and monopoles throughout  the simulation, map their history, and get an estimate on the velocity.  Table~\ref{veldirect} shows the values of the network velocities for segments and monopoles in the radiation and matter eras, for all values of $\beta$ under study. The velocities for radiation domination are higher than the corresponding velocities in matter domination, for all $\beta$. It is worth pointing out that there is little dependence on $\beta$ for these numbers, especially for the velocity of the segments.

The velocities can also be estimated using field-theoretical estimators of Eq. (\ref{Markestimators}), and the result can be found in Table~\ref{velfield}. The velocities obtained with both local-field estimators agree with each other, and as in the previous case, the velocities in radiation are higher than in matter. These values obtained from the field-estimators are somewhat different from  the ones obtained from the positions of the defects. In fact, the field estimators do not distinguish between strings and monopoles, and they give a  single number for the whole network. Unsurprisingly, in all cases the values  lie in between the velocities  of strings and monopoles.

Contrary to the velocities obtained from the positions, the   values in Table~\ref{velfield} show a trend with $\beta$. One cannot but speculate that this difference is the result of several factors in the simulation box: the ratios of densities of strings and monopoles depend on $\beta$ (for lower $\beta$ strings are longer), the tension of the strings (for lower $\beta$ strings are lighter), the tendency of strings to collapse or merge... Moreover, it has been noted before that the estimation of defect velocities using directly the position of the defects underestimates the velocities \cite{Moore:2001px,Lopez-Eiguren:2016jsy,Hindmarsh:2016dha}, and the reason for this is not completely clear.

Even though there are some discrepancies in the values of the velocities, one huge advantage of being able to follow the positions of the strings is that we can obtain the history and the velocities for individual segments.  Figures~\ref{fig:point-velocityf} and \ref{fig:point-velocitym} show the scatter plot of segment length versus segment velocity, for cases when the segments do not interact with other segments in the following time step, and for cases where the segments merge with other segments, respectively. 

It was somewhat surprising to see that the values of the velocities where scattered around a central value for all values of the length of the strings. In other words, we found no correlation between the length and the velocity of the segments. This is a hint that the VOS effective model may need to be revisited to take this into account, also in view on the recent revision of the VOS model for monopoles in \cite{Sousa:2017wvx}. Moreover,  one could have wondered whether it is more preferable for short segments to collapse and for longer ones to merge. These plots show that mergings happen for all segment sizes, not only for longer ones.

The histograms where the length of the strings per bin are depicted (Figure~\ref{histvelseg} bottom) show that actually most of the string length is not in the shortest segments; there is more length in the second and/or third bins. This histogram also shows more clearly that there are very long segments which do merge with other ones. This interesting behaviour is also shown in the family tree-like figure Fig.~\ref{b04family}. There we can see many mergings at the beginning of the simulation, then some quiet period, before mergings happen also towards the end of the simulation. The lengths of the segments that take part in mergings are very varied, and there is no correlation between segment size and their future behaviour.

Monopole velocities show, however, that segments that are going to collapse do it generally at somewhat higher than average velocities. The distribution is not so clearly Gaussian; there is more noise, and some monopoles are very fast. One has to be cautious here and remember all the caveats we have to overcome to obtain velocities of segments. The instances where monopole velocities were very close to unity (and sometimes seemingly faster than that) are due to the merging/collapsing of segments, and the shortcomings of our algorithm or interpretation of those events.  

One of the  most challenging aspects in this work has been, actually, the correct estimation of the velocity of segments and monopoles close to mergings. The na\"ive estimators of velocity gave instances of extremely high velocities. After we corrected for the obvious  caveats (such as trying to get the velocity of a segment that had collapsed by trying to find the collapsed segment in the next time step and obviously failing) many of those very fast segments disappeared. But there were still some cases where the velocity was too high, higher than what one would naively expect from the causality of the field equations, and from what the field-estimators for velocities were showing.

 The reason for this apparently superluminal behaviour is that, close to segment mergings, if the field configuration is favourable, new string segments appear  `out of the blue'. This is not a case of a monopole moving towards an antimonopole and creating string as it evolves; it is rather a new segment appearing, and giving the illusion of a very fast movement of the monopole/segment (in some sense, the far end of the string is a new monopole altogether, and the {\it old} monopole has annihilated with the newly formed antimonopole).  This behaviour is completely consistent with causal dynamics; it is the effect of the magnetic field accreting to form the new string segment. It has been studied in two-dimensional simulations in \cite{Achucarro:1992hs,Achucarro:1997cx} and is also seen in the case of a global monopole whose cores are pinned down while letting the radial field gradients bunch into a string-like region that subsequently decays and disappears, taking the monopole with it \cite{Goldhaber:1989na,Bennett:1990xy,Achucarro:2000td}.

Some of our conclusions are expected to be relevant for the 
VOS analytic model for semilocal strings. In particular, our results  suggest that --unlike in the Abelian Higgs case, where smaller loops are faster-- the velocity of semilocal segments is relatively insensitive to the size of the segments. Also, 
if we assume that network  velocities are subluminal, the effective model has to account for the possibility of segments growing out of thin air, because otherwise the speed at which segments merge and collapse  may be underestimated.  
 
This work on the characterization of semilocal strings has highlighted even more how rich the dynamics of this model is, and how complicated the life of a semilocal network can be.

\acknowledgments

We are grateful to A. Mafalda Leite and Andr\'e Nunes for their help with the simulations during the early stages of this work. A.Ach., A.L-E. and J.U.  acknowledge support from the Basque Government (IT-979-16) and the Spanish Ministry MINECO  (FPA2015-64041-C2-1P).  A.L-E. is also supported by the Basque Government grant BFI-2012-228. The work of A.Avg. has been partly supported by the University of Nottingham though a Nottingham Research Fellowship. C.J.M. is supported by an FCT Research Professorship, contract reference IF/00064/2012, funded by FCT/MCTES (Portugal) and POPH/FSE (EC). A. Ach's work was partially supported by a grant from the Simons Foundation, by the Organization for Research in Matter (FOM), by the Netherlands Organization for Scientific Research (NWO) and the Dutch Ministry of Education, Culture and Science (OCW). This work was started in the context of project PTDC/FIS/111725/2009 (FCT, Portugal). Part of this work was undertaken on the COSMOS Shared Memory system at DAMTP, University of Cambridge operated on behalf of the STFC DiRAC HPC Facility. This equipment is funded by BIS National E-infrastructure capital grant ST/J005673/1 and STFC grants ST/H008586/1, ST/K00333X/1. It  has been possible also thanks to the computing infrastructure of the i2Basque academic network.

\bibliography{semilocal}
\end{document}